\documentclass[%
 reprint,
 amsmath,amssymb,
 aps,
prb,
floatfix,
]{revtex4-2}

\usepackage{placeins}
\usepackage{xcolor}
\usepackage[normalem]{ulem}
\usepackage{comment}
\usepackage{graphicx}
\usepackage{dcolumn}
\usepackage{bm}
\usepackage{bbm}
\usepackage{bbold}
\usepackage{multirow}
\usepackage{hyperref}
\hypersetup{colorlinks=true,breaklinks,linkcolor=blue,urlcolor=blue,citecolor=blue}
\usepackage{braket}
\usepackage{comment}
\usepackage{breakcites}
\usepackage{orcidlink}

\newcommand{\TUVienna}{\affiliation{$^a$Institute of Solid State Physics, TU Wien, 1040 Vienna, Austria}}
\newcommand{\CCQ}{\affiliation{$^b$Center for Computational Quantum Physics, Flatiron Institute, 162 5th Avenue, New York, New York 10010, USA}}
\newcommand{\SISSA}{\affiliation{$^c$International School for Advanced Studies (SISSA), via Bonomea 265, 34136 Trieste, Italy}}

\newcommand{\pdg}{{\ensuremath{\phantom{\dagger}}}}

\makeatletter
\newcommand{\equalcontributionnote}{%
  \begingroup
  \let\frontmatter@footnotemark\@gobble
  \frontmatter@footnote{These authors contributed equally to this work.}%
  \endgroup
}
\makeatother

\makeatletter
\newdimen\widetextsafecolumnwidth
\newenvironment{widetextsafe}{%
  \par\ignorespaces
  \widetextsafecolumnwidth=\columnwidth
  \onecolumngrid
  \vskip30\p@
  \noindent\hb@xt@\textwidth{%
    \hb@xt@\widetextsafecolumnwidth{\leaders\hrule\hfil\vrule\@height6\p@}%
    \hfil
  }%
  \vskip6\p@
  \prep@math@patch
}{%
  \par
  \vskip6\p@
  \noindent\hb@xt@\textwidth{%
    \hfil
    \hb@xt@\widetextsafecolumnwidth{\vrule\@depth6\p@\leaders\hrule\hfil}%
  }%
  \vskip8.5\p@
  \twocolumngrid
  \global\@ignoretrue
  \@endpetrue
}
\makeatother

\begin{document}

\title{Dynamical vertex approximation for retarded interactions: \\ Application to the Hubbard-Holstein model}

\author{E.~Moghadas$^{a,*}$ \orcidlink{0009-0001-9441-2123}}
\author{S.~Giuli$^{b,c,*}$ \orcidlink{0009-0004-7341-3655}}
\author{M.~Capone$^{c}$ \orcidlink{0000-0002-9811-5089}}
\author{A.~Toschi$^{a}$ \orcidlink{0000-0001-5669-3377}}
\equalcontributionnote

\TUVienna
\CCQ
\SISSA

\date{\today}

\begin{abstract}
We extend the dynamical vertex approximation (D$\Gamma$A), in its most commonly used ladder-version, to systems with retarded electronic interactions. Such an extended scheme, which we name "dynamical-$U$ D$\Gamma$A", allows us to study strongly interacting electrons coupled to bosons including non-local spatial correlations on all length scales on top of the purely local ones captured by dynamical mean-field theory. We demonstrate the applicability of the method in the specific case of the coupling to a single dispersionless phononic mode, by studying the Hubbard-Holstein model on the square and cubic lattices, focusing on regimes of sizable Hubbard interactions where strong electronic correlations compete with phonon-mediated interactions. As illustrative examples, we consider the antiferromagnetic phase in the three-dimensional model and the $d$-wave superconductor in two dimensions, analyzing the effect of the electron-phonon coupling, including its frequency dependence, on both instabilities.

\end{abstract}

\maketitle

\section{Introduction}

Strongly correlated electron systems and their accurate theoretical description remain among the most compelling challenges in condensed-matter physics. Since its introduction, the Hubbard model \cite{Hubbard1963}, with its purely instantaneous and on-site electronic interaction, has become a paradigmatic framework for investigating correlation-driven phenomena \cite{imada1998}, including Mott physics, magnetism, pseudogap behavior, unconventional superconductivity, and quantum criticality. Hubbard-type models also provide the basis for describing a broad class of correlated materials, particularly those with partially filled $d$- or $f$-electron shells. The low-energy many-electron Hamiltonians for these systems are commonly derived through \textit{downfolding} procedures \cite{aryasetiawan2004, honerkamp2012, werner2016, aryasetiawan2022}, in which high-energy degrees of freedom are integrated out and their screening effects are incorporated into an effective electron-electron interaction, often approximated by a static Hubbard parameter $U$. However, due to the different screening processes, the effective electronic interaction encountered in real materials is not an instantaneous interaction. As a matter of fact, one of the most common methods for quantitatively evaluating the effective interaction in these low-energy models, the constrained random-phase approximation (cRPA) \cite{aryasetiawan2004, aryasetiawan2006, aryasetiawan2008} always yields a dynamical interaction $U(\omega)$. This can often have small characteristic energy scales resulting in a pronounced frequency-dependence, which, when replaced by an effective static interaction, might result in drastically overestimated screening effects \cite{amadon2014, shinaoka2015crpa, honerkamp2018, pauli2025}.

Assessing the effects of dynamical screening in these models requires a theoretical framework capable of treating the underlying strong electronic correlations. Among the methods developed to address this challenge, dynamical mean-field theory (DMFT) has emerged as a widely used approach owing to its success in describing local correlation effects. DMFT maps the lattice problem onto a quantum impurity embedded in a self-consistent bath \cite{georges1996}. Its core approximation is based on a local electronic self-energy but with a rich frequency dependence that allows us to retain the full temporal dynamics of local correlations while neglecting spatial correlations beyond the impurity. 
The advent of modern continuous-time quantum Monte Carlo (CT-QMC) impurity solvers \cite{gull2011continuous}, particularly in the hybridization-expansion formulation, \cite{werner2006hybridization, werner2006continuous} has facilitated the inclusion of retarded density-density interactions within the DMFT approach without a substantial increase in computational cost \cite{werner2007efficient, werner2010dynamical}.

This development has enabled systematic \textit{ab initio} studies of screening effects, since first-principles downfolding methods generally produce a frequency-dependent interaction $U(\omega)$. Such interactions can be incorporated, for instance, within density functional theory (DFT) + DMFT \cite{held2007rev, kotliar2006} or GW + DMFT \cite{boehnke2016, tomzcak2017, choi2016}. Retarded interactions also arise within extended DMFT (EDMFT) \cite{sengupta1995, si1996, smith2000, chitra2000}, where nonlocal Coulomb interactions are mapped onto a self-consistent dynamical impurity interaction. Building on this framework, GW and EDMFT have been combined in the GW+EDMFT \cite{sun2002, biermann2003, yeh2026} formalism  to determine the screening self-consistently through a bosonic bath, providing a unified treatment of local vertex corrections and nonlocal charge fluctuations at the DMFT level. What is, however, neglected in all of the above mentioned studies, at least beyond the GW level, is a systematic treatment of non-local correlations on top of DMFT.

At the same time, retarded interactions can result from integrating out bosonic degrees of freedom such as local phonons, which generate a retarded attractive contribution to the electronic interaction. Electronic screening and phonon-mediated interactions can therefore be described within the same general framework of a frequency-dependent effective interaction. In this respect, the minimal model that treats electronic and lattice degrees of freedom on equal footing is the Hubbard-Holstein model. Despite its local interactions and simple form, the competition between the Hubbard repulsion and electron-phonon (el-ph) coupling gives rise to a rich range of phenomena and non-trivial physics \cite{Mahan93, grilli1994, pietronero1995, kulic2001, arrigoni2003, capone2004, cappelluti2004, capone2010rev,nowadnick2012,johnston2013,murakami2013,mendl2017,costa2020,karakuzu2022, al-eryani2025, coulter2026origins}. In recent years, DMFT has played a central role in the study of the Hubbard-Holstein model, allowing the interplay between local Coulomb repulsion and phonon-mediated attraction to be treated nonperturbatively. For instance, DMFT studies have shown that the influence of phonons on electronic quasiparticles depends sensitively on the Hubbard interaction, characteristic phonon frequency and filling \cite{sangiovanni2006afm,sangiovanni2006dynamics,capone2010rev}. Near a Mott transition, phonon signatures in the spectral function can be more pronounced at higher electronic energies than in the low-energy Fermi-liquid regime \cite{sangiovanni2005,coulter2026}. Conversely, electronic correlations can substantially renormalize the effective el-ph coupling \cite{abramovitch2025,moghadas2026} and modify phonon lifetimes \cite{abramovitch2025}. Yet, in the above mentioned studies for the Hubbard-Holstein model, the effect of non-local correlations on all length scales beyond the DMFT level has not been considered.

 In recent years, two principal routes beyond DMFT have emerged, leading on one hand to cluster-extensions \cite{maier2005rev}, which include correlations exactly within the size of the finite cluster, and diagrammatic methods \cite{rohringer2018diagrammatic} on the other, that are able to capture spatial fluctuations on all length scales, albeit in an approximate manner. Diagrammatic approaches exploit the local one- and two-particle correlation functions of the auxiliary Anderson impurity model (AIM) of DMFT on top of the nonperturbative DMFT description as building blocks for constructing nonlocal corrections. Prominent examples include the dual-fermion method \cite{rubtsov2008} and the dynamical vertex approximation (D$\Gamma$A) \cite{toschi2007}. Within D$\Gamma$A, the irreducible vertices are approximated by their local impurity counterparts, and thus the nonlocal correlations beyond DMFT are generated through the parquet or the Bethe-Salpeter equation (BSE). The latter choice corresponds to the computationally more efficient ladder formulation of the approach \cite{rohringer2018diagrammatic}. For the sake of clarity, we should emphasize here that the term "dynamical" in the acronym D$\Gamma$A refers to the full frequency dependence of the local impurity vertex of DMFT. Its application so far has only considered an instantaneous bare interaction. 
 
 Diagrammatic approaches that already provide a common framework for combining dynamical screening with nonlocal collective fluctuations include the dual-boson, TRILEX, and D-TRILEX formalisms \cite{rubtsov2012,vanloon2014,ayral2015,stepanov2016,stepanov2019,stepanov2021,vandelli2022}. In this work, we provide a complementary route to dynamical screening, by extending the formalism of ladder-D$\Gamma$A to the "dynamical-$U$ D$\Gamma$A" scheme designed to include retarded interactions, consistently accounting for the frequency dependence of the bare interaction in the Bethe-Salpeter equations, the Schwinger-Dyson equation (SDE), and the Moriya-$\lambda$ corrections \cite{katanin2009comparing}. We then apply our dynamical-$U$ D$\Gamma$A approach \cite{moghadas_ldga} to the Hubbard-Holstein model, for which our method is well suited, as integrating out the Holstein phonons generates a retarded attractive contribution to the electronic interaction. 

In this context, we investigate how the strength and characteristic frequency of the el-ph coupling modify the antiferromagnetic phase diagram of the three-dimensional model, both at and away from half-filling, and analyze their effects on magnetic fluctuations, the quantum-critical regime and the nonlocal electronic self-energy. After discussing this test-bed example, we also consider a relevant 2D case. In particular, here, in two-dimensions, we study the $d$-wave superconducting instability by disentangling phonon-induced changes in the pairing interaction from their feedback on quasiparticle coherence and scattering, thereby determining how their competition affects the superconducting transition temperature.

The remainder of this paper is organized as follows. In Sec.~\ref{sec:dga}, we introduce the Hubbard--Holstein model and derive the ladder-D$\Gamma$A formalism for retarded interactions, including the corresponding Moriya-$\lambda$ corrections. In Sec.~\ref{sec:res}, we apply the method to antiferromagnetism in the three-dimensional model and to $d$-wave superconductivity in two dimensions. Finally, Sec.~\ref{sec:sum} summarizes our main findings and discusses future extensions. Additional numerical and methodological details are provided in the appendices.

\section{D$\Gamma$A algorithm for retarded interactions}\label{sec:dga}
\subsection{Model}
The starting point of our considerations is a generic lattice model with dispersion $\epsilon_{\mathbf{k}}$ and a local instantaneous Hubbard interaction
$U$ coupled to a bosonic bath,
\begin{equation}
    \begin{aligned}
    H={}&\sum_{\mathbf{k}\sigma}
        (\varepsilon_{\mathbf{k}}-\mu)
        c_{\mathbf{k}\sigma}^{\dagger}c^{\pdg}_{\mathbf{k}\sigma}
        +U\sum_i n_{i\uparrow}n_{i\downarrow} \\
      &+\sum_{i\alpha}\Omega_\alpha
        b_{i\alpha}^{\dagger}b^{\pdg}_{i\alpha}
        +\sum_{i\alpha}g_\alpha
        n_i(b_{i\alpha}^{\dagger}+b^{\pdg}_{i\alpha}).
    \end{aligned}
\end{equation}
where $c_{\mathbf{k}\sigma}^{\dagger}/c_{\mathbf{k}\sigma}^{\pdg}$ denote the fermionic and $b_{i,\alpha}^{\dagger}/b_{i,\alpha}^{\pdg}$ the bosonic creation and annihilation operators. The chemical potential $\mu$ controls the filling of the system given by $n_i = \sum_{\sigma} c^\dagger_{i \sigma} c^{\pdg}_{i \sigma}$. The bosons, with a characteristic frequency of $\Omega_{\alpha}$ couple locally via the coupling $g_{\alpha}$ to the on-site electronic density, where the index $\alpha$ labels otherwise arbitrary bosonic modes. Integrating the bosons out gives the retarded density-density interaction
\begin{equation}\label{equ:V_bath}
    V(\omega)
    =-\sum_\alpha
    \frac{2g_\alpha^2\Omega_\alpha}{\omega^2+\Omega_\alpha^2}.
\end{equation}
The retarded interaction in Eq.~\eqref{equ:V_bath} is not restricted to a coupling of the electrons to the physical lattice vibrations. More generally, a frequency-dependent local interaction naturally arises in low-energy descriptions of correlated materials as a consequence of dynamical electronic screening, which is particularly relevant for modern \textit{ab initio} approaches combining many-body perturbation theory with DMFT \cite{boehnke2016, tomzcak2017, choi2016, sun2002, biermann2003, yeh2026}. The emerging frequency-dependence can be fit to a pole-expansion according to Eq.~\eqref{equ:V_bath} with a restricted number of bosonic modes \cite{GiovannettiPRB2014} and efficiently simulated in quantum-embedding methods with a discretized impurity-bath such as exact-diagonalization \cite{capone2010rev, amaricci2022CPC, Crippa2025SPCa} or the ghost-rotationally-invariant slave-boson formalism \cite{giuli2026}. Impurity solvers based on CT-QMC, however, enable the inclusion of a $U(\omega)$ with arbitrary frequency-dependence at negligible additional computational cost \cite{werner2010dynamical}.

Without loss of generality for the subsequent formal developments, in the following we use the Hubbard-Holstein model as a representative realization of this interaction, which can be efficiently computed within DMFT \cite{werner2007efficient, werner2010dynamical} and used as a starting point of our D$\Gamma$A method.
This model extends the generic Hubbard model by an additional local Einstein mode, given by a coupling $g_{\alpha}\equiv g$ and a dispersionless characteristic frequency $\omega_0$. In particular, we consider the model on the hypercubic lattice with only nearest-neighbor hopping.
\begin{equation}\label{equ:hh_hamiltonian}
    \begin{aligned}
    H={}&-t\sum_{\langle ij\rangle,\sigma}
        \left(c_{i\sigma}^{\dagger}c^{\pdg}_{j\sigma}+\mathrm{H.c.}\right)
        -\mu\sum_i n_i+U\sum_i n_{i\uparrow}n_{i\downarrow} \\
      &+\omega_0\sum_i b_i^{\dagger}b^{\pdg}_i
        +g\sum_in_i(b_i^{\dagger}+b^{\pdg}_i),
    \end{aligned}
\end{equation}
Here, $\langle ij\rangle$ denotes a nearest-neighbor bond. Throughout this paper we restrict ourselves to the cases of $d=2$ and $d=3$ and use $D=2\sqrt{2d}\,t$ as our units of energy, which fixes the standard deviation of the non-interacting density of states (DOS) to $\sigma=0.5$ in all dimensions and allows for better comparability between different lattice geometries \cite{Bulla2000}.

In the case of a single Holstein phonon, Eq.~\eqref{equ:V_bath} reduces to
\begin{equation}\label{equ:V_hh}
    V(\omega)
    =-\frac{2g^2\omega_0}{\omega^2+\omega_0^2}
    =-\lambda \frac{\omega_0^2}{\omega^2+\omega_0^2}.
\end{equation}
with the el-ph coupling strength defined as $\lambda\equiv2g^2/\omega_0$ and thus, $V(0)=-\lambda$. In the adiabatic regime, where $\omega_0$ is small compared with electronic energy scales, the phonon-mediated attraction is strongly retarded. Exactly in the limit where $\omega_0=0$ at fixed $\lambda$, only the zero-frequency component remains, $V(\omega)=-\lambda\delta(\omega)$, corresponding at finite temperature to a classical, static phonon field. In the antiadiabatic limit, $\omega_0\to\infty$, at fixed $\lambda$, the attraction instead becomes instantaneous yielding a Hubbard-like local interaction, $V(\omega)=V(\omega=0)=-\lambda$. All subsequent expressions depend only on $V(\omega)$ and therefore apply unchanged to a general local dynamical interaction upon replacing Eq.~\eqref{equ:V_hh} by the corresponding
frequency dependence.

With an effective interaction given by $\widetilde{U}(\omega) = U + V(\omega)$ one can then define a generalized spin-dependent dynamic interaction as
\begin{equation}
    \widetilde{U}_{\sigma \sigma'}^{\nu \nu' \omega}
    = U\,(1-\delta_{\sigma \sigma'}) + V(\omega)
    - V(\nu-\nu')\,\delta_{\sigma \sigma'}.
\end{equation}
where $\nu_n=(2n+1)\pi/\beta$ and $\omega_m=2m\pi/\beta$ refer to the fermionic and  bosonic Matsubara frequencies respectively with $n,m\in\mathbb{Z}$ and $\beta=1/T$ the inverse temperature. This allows us to recast the interaction in the density (d) and magnetic (m) channels, $\widetilde{U}_{d/m} = \widetilde{U}_{\uparrow\uparrow} \pm \widetilde{U}_{\uparrow\downarrow}$
\begin{equation}\label{equ:gen_int_r}
    \begin{aligned}
        &{} \widetilde{U}_d^{\nu \nu' \omega} = 2\widetilde{U}(\omega)-\widetilde{U}(\nu-\nu')    \\
        &{}\widetilde{U}_m^{\nu \nu' \omega} = -\widetilde{U}(\nu-\nu')   
    \end{aligned}
\end{equation}

For later use, we also define the longitudinal parts of the bare density and
magnetic interaction, which contain only the transfer frequency
$\omega$ and exclude the transverse contributions that are functions of $(\nu-\nu')$:
\begin{equation}\label{equ:longitudinal_interactions}
    \mathcal{U}_d^{\omega}
    \equiv U+2V(\omega),
    \qquad
    \mathcal{U}_m^{\omega}
    \equiv -U.
\end{equation}
Note that in the anti-adiabatic limit, at fixed
$\lambda$, one has $\widetilde{U}(\omega)=\widetilde{U}(0)=U+V(0)$. Hence, in this limit, Eq.~\eqref{equ:gen_int_r} correctly reduces to the pure Hubbard case with
$U_d=U+V(0)$ and $U_m=-[U+V(0)]$, while the longitudinal parts, Eq.~\eqref{equ:longitudinal_interactions}, do not reproduce these values. We will comment on this aspect and its relevance for our method toward the end of this section.

\subsection{Ladder-D$\Gamma$A equations}

\begin{figure}[h!]
    \centering
    \includegraphics[width=0.95\linewidth]{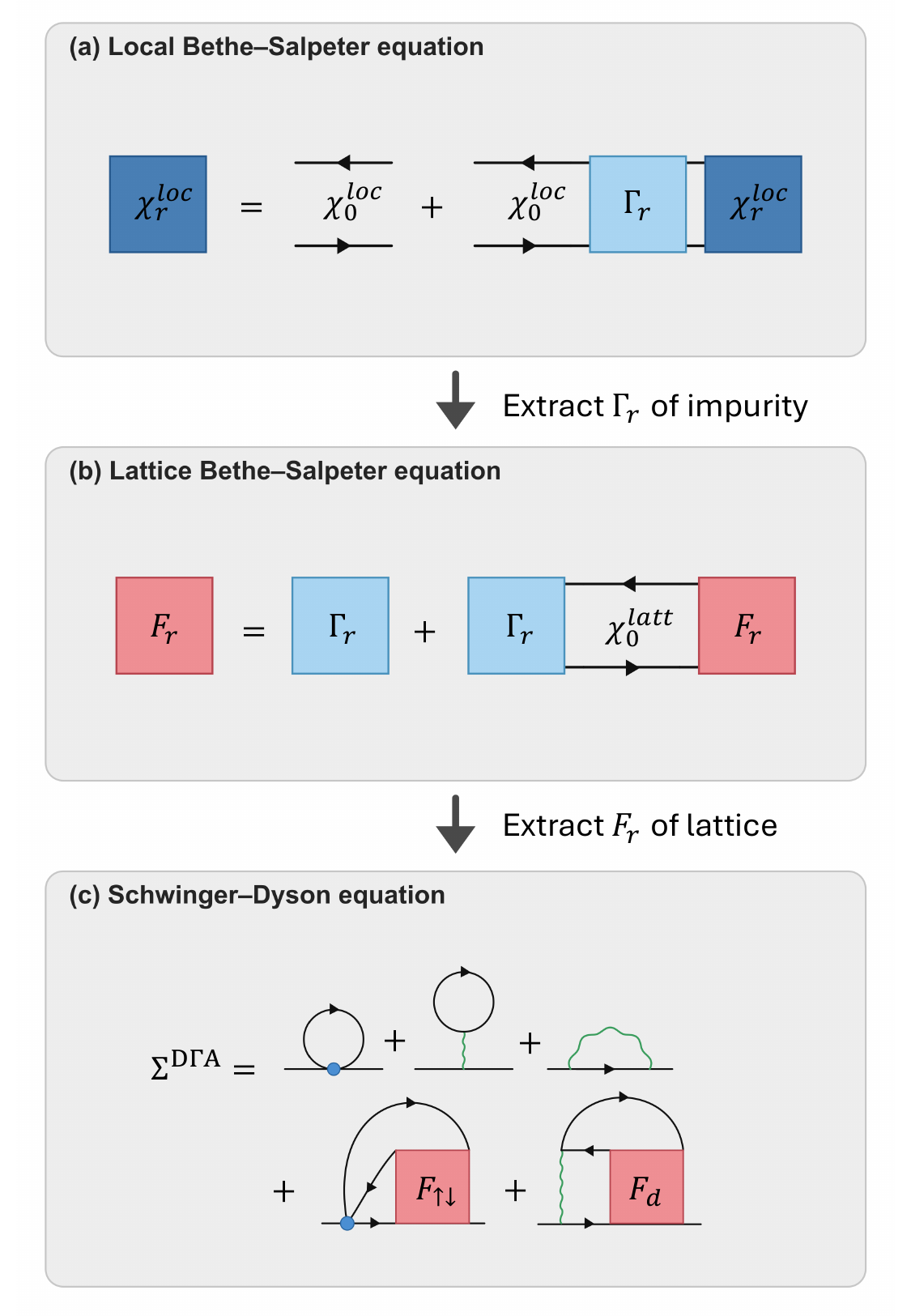}
    \caption{Schematic illustration of the dynamical-$U$ ladder-D$\Gamma$A algorithm flow. The impurity vertex $\Gamma$, obtained by inverting a local BSE (upper panel), serves as the input for the lattice ladder equation (middle panel) for the full vertex $F_{\mathbf{q}}$. This object inherits its momentum-dependence from the lattice bubble-diagrams that propagate the transfer momentum $\mathbf{q}$. In the final step the non-local correction to the DMFT self-energy is computed via the SDE (lower panel) with the DMFT ladder vertices $F_{\mathbf{q}}$ as main building blocks.}
    \label{fig:dga_flow}
\end{figure}

The fundamental difference of the present formulation of the (ladder)-D$\Gamma$A equations for retarded interactions w.r.t. the pure Hubbard case lies in the definition of the (dynamical-$U$) D$\Gamma$A SDE, also shown in the lower panel of Fig.~\ref{fig:dga_flow}, where the algorithmic flow of the method is depicted schematically. The SDE, for the exact solution, relates the self-energy to the full vertex $F$ via \cite{bickers91, janis1999, lane2026}
\begin{align}
    \Sigma(k) ={}& \mathcal{U}_d^0\frac{n}{2}
    - \beta^{-1}\sum_q V(\omega)G(k-q)
    \nonumber \\
    &- \beta^{-2} \sum_{q k'}
    \widetilde{U}(\omega)G(k')G(k+q)G(k'+q)
    F^{k k' q}_{\uparrow\downarrow}
    \nonumber \\
    &- \beta^{-2} \sum_{q k'}
    V(\omega)G(k')G(k+q)G(k'+q)
    F^{k k' q}_{\uparrow\uparrow}.
    \label{equ:sde_generic}
\end{align}
where $k=(\nu,\mathbf{k})$, $q=(\omega,\mathbf{q})$ denote four-momenta and $G=G_{\uparrow}=G_{\downarrow}$ is the SU(2)-symmetric one-particle Green's function. Due to the retarded nature of the bare interaction, in contrast to the SDE of the pure Hubbard model, here the Fock diagram (second term in the first line of Eq.~\eqref{equ:sde_generic}) as well as the $F_{\uparrow\uparrow}$ (last line of Eq.~\eqref{equ:sde_generic}) will contribute to the self-energy.

The central idea of the D$\Gamma$A formalism consists of finding suitable local approximations for building blocks of the full vertex $F$ in Eq.~\eqref{equ:sde_generic}. We recall here that $F$ contains all one-particle irreducible (1PI) two-particle diagrams and describes the full scattering amplitude between two interacting electrons in the system considered. It can be related to the generalized susceptibility \cite{rohringer2018diagrammatic}
\begin{equation}
    \chi_{\sigma\sigma'}^{k k' q} = \chi_0^{k k' q} - G_k G_{k+q} F^{k k' q}_{\sigma\sigma'}  G_{k'} G_{k'+q}
\end{equation}
where the bubble, $\chi_0^{k k' q}=-\beta G_k G_{k+q} \delta_{k k'}$, describes the independent propagation of an interacting electron-hole pair and the second term, the vertex-corrections, is essentially the connected part of the two-particle Green's function \cite{abrikosov2012methods, rohringer2014phd, rohringer2018diagrammatic}. As a second step, the diagrammatic contributions to the full vertex can be classified into two-particle reducible (2PR) or irreducible (2PI) contributions via the following parquet equation \cite{Bickers2004, rohringer2018diagrammatic}
\begin{equation}
    F_{\sigma\sigma'}^{k k' q}
    = \Lambda_{\sigma\sigma'}^{k k' q}
    + \Phi_{pp,\sigma\sigma'}^{k k' q}
    + \Phi_{ph,\sigma\sigma'}^{k k' q}
    + \Phi_{\overline{ph},\sigma\sigma'}^{k k' q}.
    \label{eq:F_parquet}
\end{equation}

Here, $\Lambda$ denotes the fully 2PI vertex, while $\Phi_r$, with
$r\in\{ph,\overline{ph},pp\}$, refer to the two-particle scattering diagrams
reducible in the particle--hole ($ph$), transverse particle--hole
($\overline{ph}$), and particle--particle ($pp$) channels, respectively
\cite{rohringer2018diagrammatic,rohringer2012local}. On the level of the
full/parquet D$\Gamma$A, one would approximate the fully irreducible vertex
$\Lambda$ by a local quantity \cite{rohringer2018diagrammatic}, obtained from
the auxiliary impurity model of DMFT \cite{georges1996}, and subsequently
obtain the nonlocal reducible objects $\Phi_r$ through a self-consistent
solution of the parquet equations for the lattice model \cite{valli2015,rohringer2018diagrammatic, kauch2020}. Due to the extremely high numerical cost of such a full parquet treatment, further approximations are often necessary. Therefore, in the ladder formulation of D$\Gamma$A \cite{toschi2007,katanin2009comparing,rohringer2018diagrammatic} one can lift the assumption of locality to the irreducible vertices $\Gamma_r$ of the dominating scattering channels.

For a chosen channel $r$, the irreducible vertex relates to the full one as
\begin{equation}
    \begin{aligned}
        F_r^{k k' q}
        &= \Gamma_r^{k k' q}+\Phi_r^{k k' q}, \\
        \Gamma_r^{k k' q}
        &= \Lambda^{k k' q}
        + \sum_{r'\ne r}\Phi_{r'}^{k k' q}, \\
    \end{aligned}
    \label{eq:F_r_bse}
\end{equation}
Thus, $\Gamma_r$ is irreducible in channel $r$ and contains the fully
irreducible vertex $\Lambda$ together with all contributions reducible in the
other channels $r'\ne r$.

The central approximation of ladder--D$\Gamma$A is to replace this generally
nonlocal channel-irreducible vertex by its local impurity counterpart $\Gamma_r^{k k' q} \simeq \Gamma_{r,\mathrm{loc}}^{\nu\nu'\omega}$. The local vertex can be obtained by inverting the impurity BSE (see upper panel of Fig~\ref{fig:dga_flow}),
\begin{equation}
    \Gamma_{r,\mathrm{loc}}^{\nu\nu'\omega}
    = \left[\chi_{r,\mathrm{loc}}^{\nu\nu'\omega}\right]^{-1}
    - \left[\chi_{0,r,\mathrm{loc}}^{\nu\nu'\omega}\right]^{-1},
\end{equation}
where $\chi^{\nu\nu'\omega}_{r,\mathrm{loc}}$ and $\chi^{\nu\omega}_{0,r,\mathrm{loc}}$ are the impurity
susceptibility and the corresponding bubble, respectively. This approximation yields a decoupling in the three channels and allows us to compute the reducible diagrams, $\Phi_r$, by means of ladder resummations in select channels, where non-local fluctuations are believed to be most relevant. In the case of a repulsive interaction, as for the pure Hubbard model, or in our particular case, where we only consider the physically more relevant regime of a dynamical interaction which satisfies $|V(\omega=0)|<U$, the dominant fluctuations arise arguably from magnetic scattering processes in the \textit{ph}-channel \cite{toschi2007,rohringer2018diagrammatic,delre2019}. Hence, in addition to considering only local diagrams in the fully irreducible vertex, the \textit{pp}-channel is approximated by its local counterpart.
\begin{equation}\label{equ:lam_phi_pp_loc}
    \Lambda_{\sigma\sigma'}^{k k' q}
    \simeq \Lambda_{\mathrm{loc},\sigma\sigma'}^{\nu\nu'\omega},
    \qquad
    \Phi_{pp,\sigma\sigma'}^{k k' q}
    \simeq \Phi_{pp,\mathrm{loc},\sigma\sigma'}^{\nu\nu'\omega}.
\end{equation}

This way, the \textit{ph} ladder-diagrams can be resummed using $\Gamma_{r,\mathrm{loc}}$ and the lattice Green's function of DMFT $G_k =\left[i\nu + \epsilon_{\mathbf{k}} - \mu + \Sigma(\nu) \right]^{-1}$, with $\Sigma(\nu)$ being the impurity self-energy. We stress that, despite the lack of momentum-dependence in the irreducible vertex $\Gamma_r$, the resulting full vertex we obtain from resumming the lattice-BSE remains nonlocal through the \textit{ph}-bubble propagating the transfer momentum $\mathbf{q}$.
\begin{equation}
    F_{r,\mathbf{q}}^{\nu\nu'\omega}
    = \Gamma_{r,\mathrm{loc}}^{\nu\nu'\omega}
    + \beta^{-1}\sum_{k_1}
    \Gamma_{r,\mathrm{loc}}^{\nu\nu_1\omega}
    G_{\mathbf{k_1}}^{\nu_1}G_{\mathbf{k_1+q}}^{\nu_1+\omega}
    F_{r,\mathbf{q}}^{\nu_1\nu'\omega}.
    \label{eq:F_r_q_ladder}
\end{equation}

The second term in the r.h.s. of Eq.~\eqref{eq:F_r_q_ladder} is what we denote as the reducible contribution to the ladder vertex. For the density and magnetic channels
$r\in\{d,m\}$ it reads
\begin{equation}\label{equ:phi_ladder}
    \Phi_{r,\sigma\sigma'}^{k k' q}\equiv \Phi_{r,\mathbf{q}}^{\nu\nu'\omega}
    \equiv F_{r,\mathbf{q}}^{\nu\nu'\omega}
    - \Gamma_{r,\mathrm{loc}}^{\nu\nu'\omega}
\end{equation}
with $F_{d/m} \equiv F_{ph,\uparrow\uparrow} \pm F_{ph,\uparrow\downarrow}$. Plugging Eqs.~\eqref{equ:lam_phi_pp_loc} and~\eqref{equ:phi_ladder} into Eq.~\eqref{eq:F_parquet} gives the equal- and opposite-spin ladder vertices \cite{toschi2007, rohringer2018diagrammatic,rohringer2014phd}:
\begin{equation}
    \begin{aligned}
    \widetilde{F}_{\uparrow\uparrow}^{k k' q}
    ={}& \frac{1}{2}\left(
        F_{d,\mathbf{q}}^{\nu\nu'\omega}
        +F_{m,\mathbf{q}}^{\nu\nu'\omega}\right) \\
    &-\frac{1}{2}
        F_{d,\mathbf{k}'-\mathbf{k}}^{\nu(\nu+\omega)(\nu'-\nu)}
      -\frac{1}{2}
        F_{m,\mathbf{k}'-\mathbf{k}}^{\nu(\nu+\omega)(\nu'-\nu)} \\
    &-\frac{1}{2}\left(
        F_{d,\mathrm{loc}}^{\nu\nu'\omega}
        +F_{m,\mathrm{loc}}^{\nu\nu'\omega}\right).
    \end{aligned}
    \label{eq:F_upup}
\end{equation}

\begin{equation}
    \begin{aligned}
    \widetilde{F}_{\uparrow\downarrow}^{k k' q}
    ={}& \frac{1}{2}\left(
        F_{d,\mathbf{q}}^{\nu\nu'\omega}
        -F_{m,\mathbf{q}}^{\nu\nu'\omega}\right) \\
    &-F_{m,\mathbf{k}'-\mathbf{k}}^{\nu(\nu+\omega)(\nu'-\nu)} \\
    &-\frac{1}{2}\left(
        F_{d,\mathrm{loc}}^{\nu\nu'\omega}
        -F_{m,\mathrm{loc}}^{\nu\nu'\omega}\right).
    \end{aligned}
    \label{eq:F_updn}
\end{equation}
where the last lines can be seen as local double counting corrections. Here we used $\widetilde{F}$ for the D$\Gamma$A vertices to differentiate between the ladder variants $F_{\mathbf{q}}$. In this way, using the expressions of Eq.~\eqref{eq:F_upup} and~\eqref{eq:F_updn} for the $F_{\sigma\sigma'}$ in Eq.~\eqref{equ:sde_generic}, the SDE can be rewritten identically as
\begin{widetextsafe}
\begin{equation}\label{equ:sde_dga}
    \begin{aligned}
    \Sigma_{\sigma}(k) ={}&
        \mathcal{U}_d^0\frac{n}{2}
        -\beta^{-1}\sum_q V(\omega)G(k-q) \\
    &+\frac{\beta^{-3}}{2}\sum_{q\nu'}
        G(k+q)\sum_{\nu'}
        \Bigl[\widetilde{U}_d^{\nu\nu'\omega}
        F_{d,\mathbf{q}}^{\nu\nu'\omega}
        +3\widetilde{U}_m^{\nu\nu'\omega}F_{m,\mathbf{q}}^{\nu\nu'\omega}
        \Bigr]\chi_{0,\mathbf{q}}^{\nu'\omega} \\
    &-\frac{\beta^{-3}}{2}\sum_{q\nu'}
        \widetilde{U}(\omega)G(k+q)
        \Bigl[2F_{d,\mathrm{loc}}^{\nu\nu'\omega}\Bigr]
        \chi_{0,\mathbf{q}}^{\nu'\omega}.
    \end{aligned}
\end{equation}
\end{widetextsafe}
where the lattice \textit{ph}-bubble is defined as
$\chi_{0,\mathbf{q}}^{\nu\omega}=-\beta\sum_{\mathbf{k}}
G_{\mathbf{k}}^{\nu}G_{\mathbf{k+q}}^{\nu+\omega}$ and
$\chi_{0,\mathrm{loc}}^{\nu\omega}=\sum_{\mathbf{q}}
\chi_{0,\mathbf{q}}^{\nu\omega}$ denotes its local counterpart. We should emphasize that the SDE of the ladder-D$\Gamma$A formalism presented here, in contrast to
Eq.~\eqref{equ:sde_generic}, will contain the generalized bare interactions
$\widetilde{U}_r^{\nu\nu'\omega}$ (cf.~second line of Eq.~\eqref{equ:sde_dga}). This is a direct consequence of rewriting the
transversal ladders in Eqs.~\eqref{eq:F_upup} and~\eqref{eq:F_updn}, which
propagate their four-momentum $\sim(k'-k)$, in the \textit{ph}-notation used for the
momentum and frequency integrals of the SDE. This in turn requires a shift of
the frequency arguments of the interaction $\widetilde{U}(\omega)$.

\subsection{Moriya-$\lambda$ corrections}\label{sec:lamcorr}

\begin{figure}
    \centering
    \includegraphics[width=1.0\linewidth]{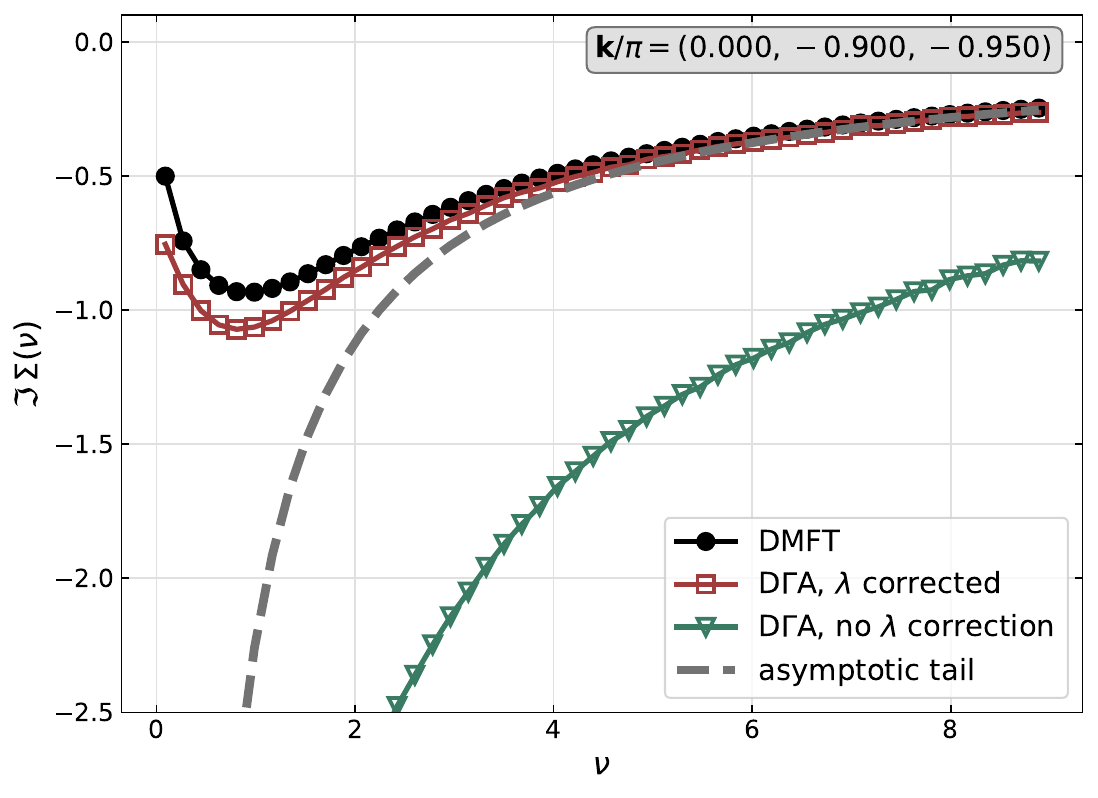}
    \caption{Comparison of $\lambda$-corrected (red) and uncorrected D$\Gamma$A (green) self-energies on the Fermi-surface with the local DMFT self-energy (black) along with the exact self-energy $1/\nu$ tail (gray dashed line). The data are obtained for the 3D Hubbard-Holstein model at $U=3$, $\lambda=1$, $\omega_0=0.1$, $n=0.9$ and $\beta=100$.}
    \label{fig:sde_asymp}
\end{figure}

It has been already extensively discussed in the previous D$\Gamma$A literature \cite{katanin2009comparing,rohringer2018diagrammatic,delre2019} that the formulation of the SDE by means of ladder diagrams leads to a spurious behavior of the self-energy asymptotics. In the following we argue that similar problems occur in the ladder-D$\Gamma$A equations for retarded interactions and how one can resolve them.

In the conventional ladder version for the repulsive-$U$ Hubbard model, the $1/i\nu$ prefactor of the D$\Gamma$A self-energy deviates from the correct $U\frac{n}{2}(1-\frac{n}{2})$ value. It can be shown \cite{rohringer2018diagrammatic,rohringer2014phd} that this is a direct consequence of the following sum rule for the physical susceptibilities $\chi_r$ associated to the Pauli principle being violated

\begin{equation}\label{equ:sum-rule_pauli}
    \frac{1}{2\beta} \sum_{q} \left(\chi_{m}^{q} + \chi_{d}^{q} \right) = \frac{n}{2}(1-\frac{n}{2})
\end{equation}

In order to fulfill the sum rule and restore the correct self-energy tail, based on the Moriya theory of itinerant magnetism \cite{Moriya1985}, one can introduce \cite{katanin2009comparing} the free parameter $\lambda_r$ in the definition of the DMFT response functions
\begin{equation}\label{equ:chi_lam_corr}
    \chi_{r,\mathbf{q}}^{\lambda_r,\omega}
    = \left[(\chi_{r,\mathbf{q}}^\omega)^{-1}
    +\lambda_r\right]^{-1}
\end{equation}
where the specific value of $\lambda_r$ is determined such that the sum-rule of Eq.~\eqref{equ:sum-rule_pauli} is fulfilled. In principle, this can be done by either correcting only the magnetic susceptibility or by correcting both channels simply by imposing that the local lattice and impurity equal-time responses are equivalent:
\begin{equation}\label{equ:loc_imp}
    \sum_{\omega, \mathbf{q}}
    \left[(\chi_{r,\mathbf{q}}^\omega)^{-1}
    +\lambda_r\right]^{-1} = \sum_{\omega} \chi_{r, IMP}^{\omega}
\end{equation}
A third way of computing the correct $\lambda_r$-parameters, as introduced in Refs.~\cite{rohringer2016impact,stobbe2022,titvinidze2025}, focuses on restoring consistency of the potential (kinetic) energy obtained via one and two-particle quantities.

An important advantage of the $\lambda$-correction is that it also fulfills the Mermin-Wagner theorem and hence yields a vanishing Néel temperature in two dimensions \cite{katanin2009comparing,rohringer2016impact,rohringer2018diagrammatic}. More generally, as mentioned in Refs.~\cite{delre2019,titvinidze2025,leusch2026} this scheme appears to transform the DMFT universality class to that of the two-particle-self-consistent (TPSC) method \cite{dare1996,vilk1997} or, equivalently, to that of the spherical symmetric ($\mathcal{O}(\infty)$ or Ka\v{c} \cite{semon2012}) model. This class of critical exponents becomes exact for an order parameter with infinite components (large $N$ limit), which can be a reasonable starting-point for the spin response associated to a three-component magnetization order-parameter. However, it is an extremely crude approximation for the charge susceptibility, which belongs to the Ising universality class ($N$=1). Consequently, in 2D the implementation of $\lambda$-corrections would lead to a vanishing transition temperature for any potentially existing charge order.

In order to discuss the appropriate $\lambda$-correction scheme for the case of retarded interactions, we first have to derive the $1/i\nu$-tail of the corresponding equation of motion, i.e., Eq.~\eqref{equ:sde_generic}. To this end we need to determine the $\nu$-independent diagrams in $F^{kk'q}$ as the $1/i\nu$ already originates from the large frequency behaviour of the Green's function depending on the four-momentum $k$, $G(k+q) \sim \frac{1}{i\nu} + \mathcal{O}(\nu^{-2})$ \cite{rohringer2014phd,rohringer2016impact}. In contrast to the pure Hubbard case, where the instantaneous interaction forbids equal-spin alignments, we need to consider both spin combinations. The large-$\nu$ asymptotics of the spin-resolved vertices can be expressed in terms of the longitudinal bare interactions $\mathcal{U}_r^\omega$ defined above. For opposite spins one obtains
\begin{align}
    F_{\uparrow\downarrow}^{k k' q}
    ={}& \sum_{k_1}\Bigl\{
        \beta\delta_{k_1k'}\widetilde{U}(\omega)
        \nonumber\\
    &\quad+UG(k_1)G(k_1+q)F_{\downarrow\downarrow}^{k_1 k' q}
        \nonumber\\
    &\quad+V(\omega)G(k_1)G(k_1+q)
        \sum_{\sigma}F_{\sigma\downarrow}^{k_1 k' q}\Bigr\}
        +\mathcal{O}(\nu^{-1})
        \nonumber\\
    ={}& \frac{1}{2}\sum_{k_1}\Bigl\{
        \beta\delta_{k_1k'}
        [\mathcal{U}_d^\omega-\mathcal{U}_m^\omega] 
        \nonumber\\
    &\quad+\chi_0^{k_1q}\bigl[
        \mathcal{U}_d^\omega F_d^{k_1 k' q}
        -\mathcal{U}_m^\omega F_m^{k_1 k' q}
        \bigr]\Bigr\} +\mathcal{O}(\nu^{-1})
        \nonumber\\
    ={}&-\frac{\beta}{2\chi_0^{k'q}}\sum_{k_1}\Bigl[
        \mathcal{U}_d^\omega\chi_d^{k_1 k' q}
        -\mathcal{U}_m^\omega\chi_m^{k_1 k' q}
        \Bigr] +\mathcal{O}(\nu^{-1}).
\end{align}
For equal spins, the corresponding result is
\begin{align}
    F_{\uparrow\uparrow}^{k k' q}
    ={}& \sum_{k_1}\Bigl\{
        \beta\delta_{k_1k'}V(\omega)
        \nonumber\\
    &\quad+UG(k_1)G(k_1+q)F_{\downarrow\uparrow}^{k_1 k' q}
        \nonumber\\
    &\quad+V(\omega)G(k_1)G(k_1+q)
        \sum_{\sigma}F_{\sigma\uparrow}^{k_1 k' q}\Bigr\}
        +\mathcal{O}(\nu^{-1})
        \nonumber\\
    ={}& \frac{1}{2}\sum_{k_1}\Bigl\{
        \beta\delta_{k_1k'}
        [\mathcal{U}_d^\omega+\mathcal{U}_m^\omega]
        \nonumber\\
    &\quad+\chi_0^{k_1q}\bigl[
        \mathcal{U}_d^\omega F_d^{k_1 k' q}
        +\mathcal{U}_m^\omega F_m^{k_1 k' q}
        \bigr]\Bigr\} +\mathcal{O}(\nu^{-1})
        \nonumber\\
    ={}&-\frac{\beta}{2\chi_0^{k'q}}\sum_{k_1}\Bigl[
        \mathcal{U}_d^\omega\chi_d^{k_1 k' q}
        +\mathcal{U}_m^\omega\chi_m^{k_1 k' q}
        \Bigr] +\mathcal{O}(\nu^{-1}).
\end{align}

Consequently, using these results in Eq.~\eqref{equ:sde_generic} leads to the following high-frequency behavior of the self-energy
\begin{align}
    \Sigma(k) ={}& \mathcal{U}_d^0\frac{n}{2}
        -\frac{1}{i\nu}\left[\beta^{-1}\sum_q V(\omega)\right]\nonumber\\
    &+\frac{1}{i\nu}\left[
        \beta^{-1}\sum_q
        \bigl(\mathcal{U}_d^\omega\bigr)^2
        \chi_{d}^q\right]
        \nonumber\\
    &+\frac{U}{2i\nu}\left[
        \beta^{-1}\sum_q\left(
        \mathcal{U}_d^\omega\chi_{d}^q
        +\mathcal{U}_m^\omega\chi_{m}^q\right)\right]
        +\mathcal{O}(\nu^{-2}).
\end{align}
Evidently, this expression cannot be immediately connected to a specific sum-rule as in Eq.~\eqref{equ:sum-rule_pauli}. Hence, the most straightforward way to restore the correct tail-behavior is by employing $\lambda_r$ parameters in both channels and imposing Eq.~\eqref{equ:loc_imp} that the $q$-integrated lattice susceptibilities correspond to their impurity counterparts. Alternatively, one could of course only correct the magnetic channel by fulfilling the Pauli sum-rule of Eq.~\eqref{equ:sum-rule_pauli} similar to conventional ladder-D$\Gamma$A at the cost of not retaining the correct self-energy asymptotics.
We refer to Appendix~\ref{app:lambda_correction} for the $\lambda$-correction scheme that follows the method of Ref.~\cite{stobbe2022} adapted for retarded interactions. In this way, the $\lambda_r$ parameters can be computed via coupled root-finding equations such that the potential energy obtained via one and two-particle quantities is consistently determined. The comparison and the role of the different correction methods for the energetics of systems with retarded interactions is however beyond the scope of the current study.

Lastly, following the steps in \cite{katanin2009comparing}, we need to reformulate the expression of the full ladder vertices $F_{\mathbf{q}}$ in Eq.~\eqref{eq:F_r_q_ladder} in terms of the $\lambda$-corrected physical susceptibilities. We define the physical response functions and the three-leg vertices via the generalized susceptibilities
\begin{equation}\label{equ:chi_gamma_q}
    \begin{aligned}
    \chi_{r,\mathbf{q}}^\omega
    &\equiv \beta^{-2}\sum_{\nu\nu'}
        \chi_{r,\mathbf{q}}^{\nu\nu'\omega}, \\
    \gamma_{r,\mathbf{q}}^{\nu\omega}
    &\equiv (\chi_{0,\mathbf{q}}^{\nu\omega})^{-1}
        \frac{\beta^{-1}\sum_{\nu_1}
        \chi_{r,\mathbf{q}}^{\nu\nu_1\omega}}
        {1-\mathcal{U}_r^\omega\chi_{r,\mathbf{q}}^\omega},
    \end{aligned}
\end{equation}
which allows us to restrict the implementation of the Moriya correction to the two-point susceptibility. The $\gamma_r$ vertices already contain the correct asymptotics and are only moderately affected by non-local correlations \cite{rohringer2016impact} and hence remain unchanged. Using the definition of the $\lambda$-corrected susceptibilities in Eq.~\eqref{equ:chi_lam_corr} we can define an associated \textit{screened interaction}
\begin{equation}\label{equ:screened_W}
    W_{r,\mathbf{q}}^{\lambda_r,\omega}\equiv \mathcal{U}_r^\omega
        \left[1-\mathcal{U}_r^\omega
        \chi_{r,\mathbf{q}}^{\lambda_r,\omega}\right]
\end{equation}
and an interaction-irreducible generalized susceptibility
\begin{equation}\label{equ:chi_star}
    \chi^{*,\nu\nu'\omega}_{r,\mathbf{q}} = \chi_{r,\mathbf{q}}^{\nu\nu'\omega} + \chi_{0,\mathbf{q}}^{\nu\omega}\gamma_{r,\mathbf{q}}^{\nu\omega}W_{r,\mathbf{q}}^{\lambda_r,\omega}\gamma_{r,\mathbf{q}}^{\nu'\omega}\chi_{0,\mathbf{q}}^{\nu'\omega}.
\end{equation}
Using the above two definitions allows us to express the \textit{$\lambda$-corrected} full vertex according to
\begin{equation}
    F_{r,\mathbf{q}}^{\lambda_r,\nu\nu'\omega}=\mathcal{R}_{r,\mathbf{q}}^{\nu\nu'\omega}
        +\gamma_{r,\mathbf{q}}^{\nu\omega}
        W_{r,\mathbf{q}}^{\lambda_r,\omega}
        \gamma_{r,\mathbf{q}}^{\nu'\omega},
\end{equation}
\begin{equation}
    \mathcal{R}_{r,\mathbf{q}}^{\nu\nu'\omega} = \left(\chi_{0,\mathbf{q}}^{\nu\omega}\right)^{-1}\,\left[\beta\delta_{\nu\nu'} - \chi^{*,\nu\nu'\omega}_{r,\mathbf{q}}\left(\chi_{0,\mathbf{q}}^{\nu'\omega}\right)^{-1}\right]
\end{equation}
Here, $\mathcal{R}_r$ denotes the residual contribution that is irreducible
with respect to cutting a single longitudinal interaction line.
In the high-frequency sector considered above it is sub-leading, so the vertex-asymptotics here is
controlled by the product of two three-leg vertices and the Moriya-corrected \textit{screened interaction} $W_r^{\lambda_r}$.

In Fig.~\ref{fig:sde_asymp} we demonstrate how the proposed scheme successfully restores the correct asymptotics of the self-energy. We compare the imaginary part of the DMFT (bold black line) with $\lambda$-corrected and uncorrected ladder-D$\Gamma$A results for the Hubbard-Holstein model in 3D computed at $\beta=100, U=3, \lambda=1, \omega_0=0.1$ at a filling of $n=0.9$.

One caveat of the presented $\lambda$-corrections here originates from the longitudinal bare interactions defined in Eq.~\eqref{equ:longitudinal_interactions}. As pointed out already before, these can lead to erroneous results of the ladder-D$\Gamma$A approaching the antiadiabatic limit $V(\omega) = V$ or more specifically $\omega_0 \to \infty$ for the Hubbard-Holstein model. We present a more detailed discussion on the anti-adiabatic limit and quantify the error caused by increasing the frequency $\omega_0$ in Appendix~\ref{app:comp_details}. As discussed there, this choice of longitudinal bare interactions leads to an increasing error as $\omega_0\to\infty$. This does not affect the purpose of the present one-shot formulation, which is to study retarded interactions at finite phonon frequencies. The instantaneous limit is better treated using the conventional ladder-D$\Gamma$A with a frequency-independent interaction. Instead, if one is interested in a controlled study across the full range from adiabatic to antiadiabatic frequencies one should employ the self-consistent D$\Gamma$A \cite{kaufmann2021}, which does not require the present Moriya-$\lambda$ corrections.

\section{Application to the Hubbard Holstein model}\label{sec:res}

We apply the extended ladder-D$\Gamma$A formalism for dynamical interactions to the finite-dimensional Hubbard-Holstein model, focusing on regimes where the sizable Hubbard repulsion favors magnetism or unconventional superconductivity. This allows us to examine how phonon-induced screening and scattering modify these instabilities once nonlocal correlations are included. Although spin fluctuations dominate the antiferromagnetic regime, charge fluctuations may also contribute to superconducting pairing. We therefore treat the two channels according to the physics of each application and therefore specify the appropriate choice of the Moriya-$\lambda$ correction in the corresponding sections.

We first consider the three dimensional model, where long-range  antiferromagnetism is possible and present for the pure Hubbard model, focusing on the dependence of the N\'eel transition on doping and phonon dynamics, and subsequently analyze the evolution of $d$-wave superconductivity with el-ph coupling in two dimensions. Throughout this section, results at finite phonon frequencies are compared with the anti-adiabatic limit, in which the Hubbard-Holstein model reduces to a purely electronic Hubbard model with an effective interaction $U_{\mathrm{eff}}=U-\lambda$, providing a reference to established DMFT \cite{ciuchi1997, ciuchi1999, capone2003, capone2004, capone2006, capone2006competition, capone2006prb, sangiovanni2005, sangiovanni2006afm, sangiovanni2006dynamics, abramovitch2025, coulter2026} and conventional D$\Gamma$A \cite{rohringer2018diagrammatic, kitatani2019, delre2019, kitatani2022, stobbe2022, leusch2026} results. The details on the DMFT and D$\Gamma$A computations are laid out in Appendix~\ref{app:comp_details}. Appendix~\ref{app:asymp} describes how the known high-frequency behavior of the DMFT vertices is used to evaluate contributions outside the explicitly sampled Matsubara-frequency box. This reduces the numerical cost of the BSE while controlling truncation errors in the calculated susceptibilities.

\subsection{Antiferromagnetism in 3D}\label{sec:afm}

To place the current D$\Gamma$A study of the magnetic phase transition in the context of the existing literature we briefly review the $T-U$ as well as $T-n$ phase-diagram of the 3D Hubbard model as obtained in Refs.~\cite{rohringer2011critical,schäfer2017,stobbe2022, fus2026}. In the half-filled system, DMFT as well as ladder-D$\Gamma$A yield a finite transition temperature to the ordered state with a maximum around $U=2$ (in units of $D=2\sqrt{2d}\,t$) \cite{rohringer2011critical,stobbe2022,leusch2026}. In DMFT, this antiferromagnetic dome arises because $T_N$ first increases with $U$ in the weak-coupling regime but decreases at strong coupling, where magnetic order is governed by the superexchange scale $J\propto t^2/U$. Including spatial correlations generally leads to a suppressed Néel temperature $T_N$ for all interaction values, albeit with a comparably milder effect in the weak coupling regime \cite{stobbe2022}. At the same time, doping the model out of half-filling has the similar effect of reducing the magnetic transition temperature until it evidently reaches zero, terminating in a quantum-critical-point (QCP). Notably, around this region, the presence of Kohn-anomalies in the underlying Fermi-surface defines the quantum-critical behavior above the QCP and shapes the entire phase-diagram around it \cite{schäfer2017,fus2026}. This phenomenon is directly tied to the Fermi-surface properties of the model and is observed in ladder-D$\Gamma$A, DMFT and even the random-phase-approximation (RPA) alike. This originates from the fact that in the realm of the single-orbital model considered here for all of the mentioned methods, the self-energies used to dress the Green's functions in the physical spin susceptibility \cite{vilardi2018} functions do not contain nonlocal self-energy corrections which could alter the DMFT/RPA metallic state and hence the underlying Fermi-surface geometry. 

In our work, we will first analyze the dependence of $T_N$ on the el-ph coupling $\lambda$ at half filling and subsequently examine its evolution upon doping. Throughout our study of antiferromagnetism, we consider three characteristic phonon frequencies, $\omega_0=\{ 0.1, 1.0, \infty \}$ in order to highlight the effects of retardation. 
Lastly, we opt for the $\lambda$-correction scheme as described in Sec.~\ref{sec:lamcorr}, where both channels, spin and charge, are corrected separately, ensuring the exact asymptotic behavior of the self-energy.

\subsubsection{Half-filled Model}

\begin{figure}
    \centering
    \includegraphics[width=1.0\linewidth]{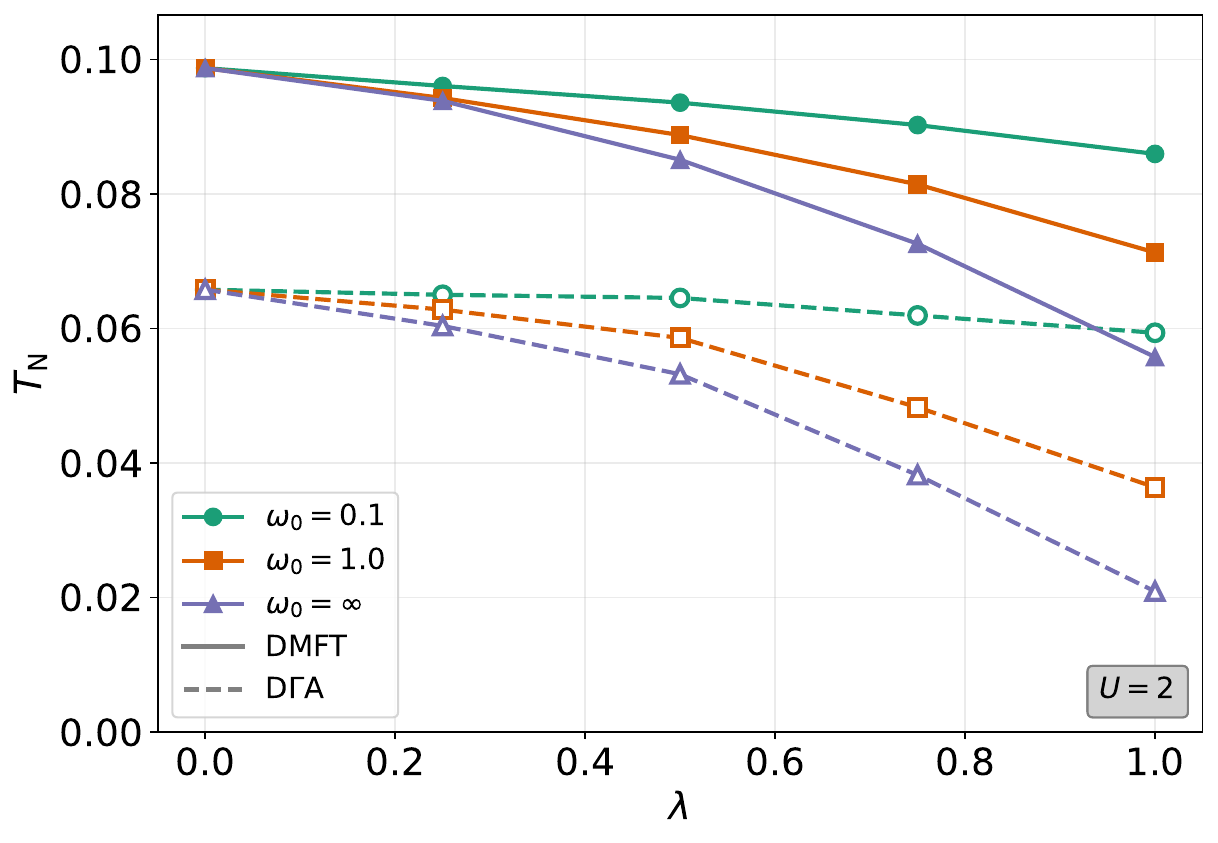}
    \caption{DMFT and ladder-D$\Gamma$A results for the Néel temperature $T_N$ of the HHM on the cubic lattice as a function of $\lambda$ and the three phonon frequencies $\omega_0=\{0.1,1,\infty \}$. The Hubbard interaction is fixed to $U=2$.}
    \label{fig:hf_TN}
\end{figure}

We begin by considering the half-filled Hubbard-Holstein model in three dimensions. For each of the three phonon frequencies, we fix the Hubbard interaction to $U=2$, close to the maximum of the antiferromagnetic dome, and vary the el-ph coupling $\lambda$. To determine $T_N$ within DMFT and ladder-D$\Gamma$A, we fit the temperature dependence of the static magnetic susceptibility. At half filling, DMFT exhibits the mean-field critical exponent $\gamma=1$, whereas, as discussed above, ladder D$\Gamma$A with the Moriya-$\lambda$ correction yields, in 3D, the TPSC/$\mathcal{O}(\infty)$ exponent $\gamma=2$. We therefore extract the N\'eel temperature by fitting the susceptibility at the antiferromagnetic ordering vector $\mathbf{Q}=(\pi,\pi,\pi)$ according to $\chi_m(\mathbf{Q},\omega=0)\propto(T-T_N)^{-\gamma}$. Details of the fitting procedure are provided in Appendix~\ref{app:fit}.

Fig.~\ref{fig:hf_TN} shows the resulting $\lambda$ dependence of $T_N$ at half-filling for the three phonon frequencies. For all values of $\omega_0$, DMFT (solid lines) yields a higher transition temperature than D$\Gamma$A (dashed lines), as a result of the suppression of magnetic order by spatial correlations absent in DMFT. In both approaches, $T_N$ is more robust against the el-ph coupling at smaller $\omega_0$, indicating that more adiabatic phonons screen the Hubbard interaction less efficiently. In fact, the competition between the local repulsion $U$ and the phonon-mediated attraction $\lambda$ at half filling has been studied extensively within DMFT \cite{sangiovanni2005,sangiovanni2006afm,sangiovanni2006dynamics,paci2006,capone2010rev}. As discussed in Ref.~\cite{sangiovanni2006dynamics}, the leading effect of the phonons can be represented as a partial screening of the electronic repulsion, controlled primarily by the ratio $\omega_0/U$. In the large-$U$ limit a static effective interaction was derived 
\begin{equation}\label{equ:Ueff_screening}
    U_{\mathrm{eff}} = U - \eta \, \lambda, \qquad
    \eta = \frac{2\,\omega_0/U}{1+2\,\omega_0/U}.
\end{equation}
The above equation interpolates between $\eta\to0$ in the adiabatic limit and $\eta\to1$ in the antiadiabatic limit. Consequently, the phonon-induced reduction of the effective repulsion becomes stronger as $\omega_0$ increases. This trend accounts qualitatively for the hierarchy observed in Fig.~\ref{fig:hf_TN}: $T_N$ depends only weakly on $\lambda$ for $\omega_0=0.1$, is more strongly suppressed for $\omega_0=1$, and decreases most rapidly in the antiadiabatic limit. Spatial correlations lower $T_N$ substantially at the D$\Gamma$A level while preserving, almost unaltered, the characteristic $\omega_0$-dependence.

\subsubsection{Doped Model}

\begin{figure*}
    \centering
    \includegraphics[width=1.0\linewidth]{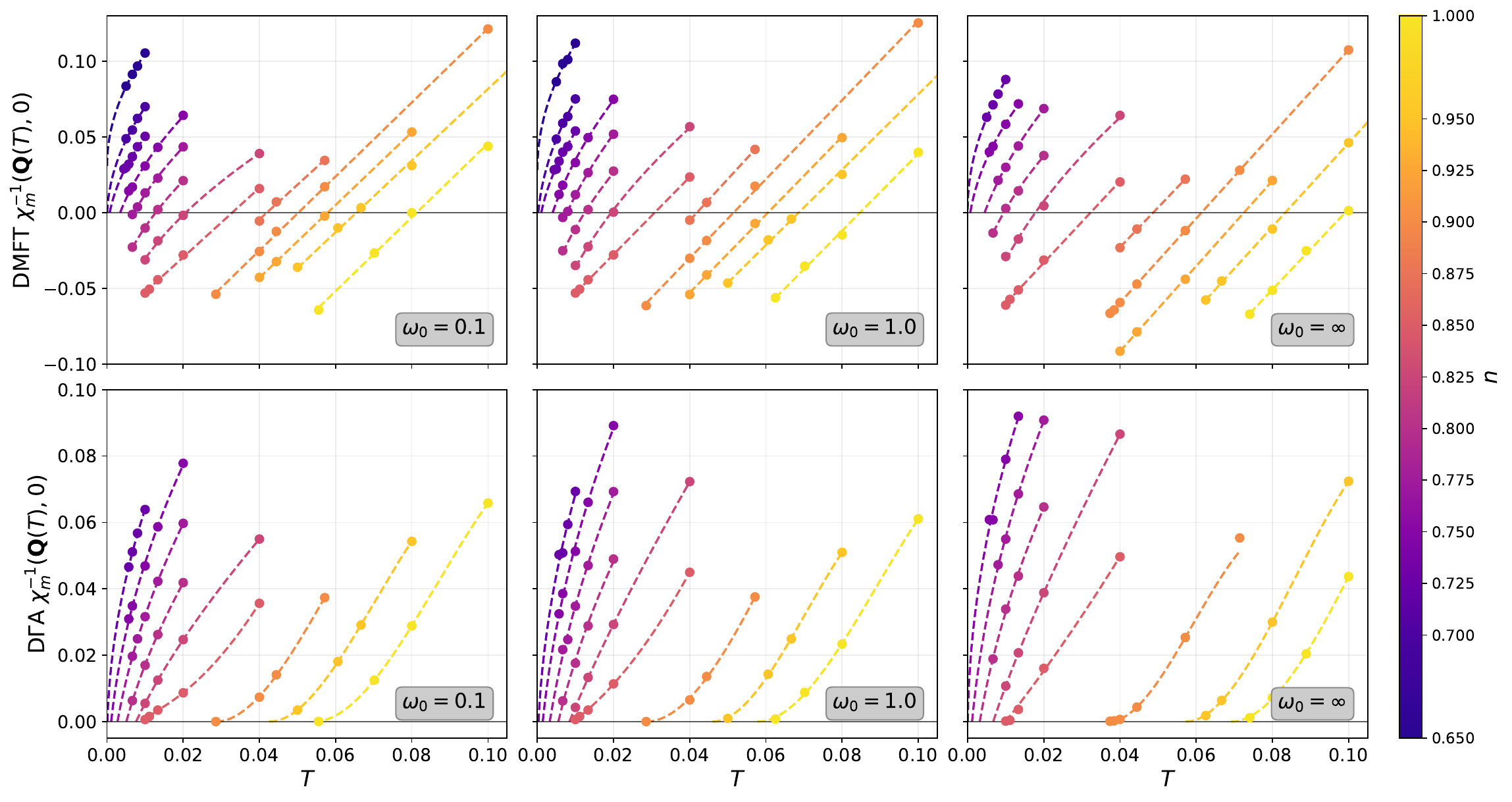}
    \caption{Temperature dependence of the DMFT (upper row) and ladder-D$\Gamma$A (lower row) static spin susceptibilities at the ordering vector $\mathbf{Q}$ for various fillings and the three phonon frequencies $\omega_0=\{0.1,1,\infty \}$. For all calculations considered here the Hubbard interaction is set to $U=3$ with the el-ph coupling being $\lambda=1$}
    \label{fig:chi_T}
\end{figure*}

\begin{figure*}
    \centering
    \includegraphics[width=1.0\linewidth]{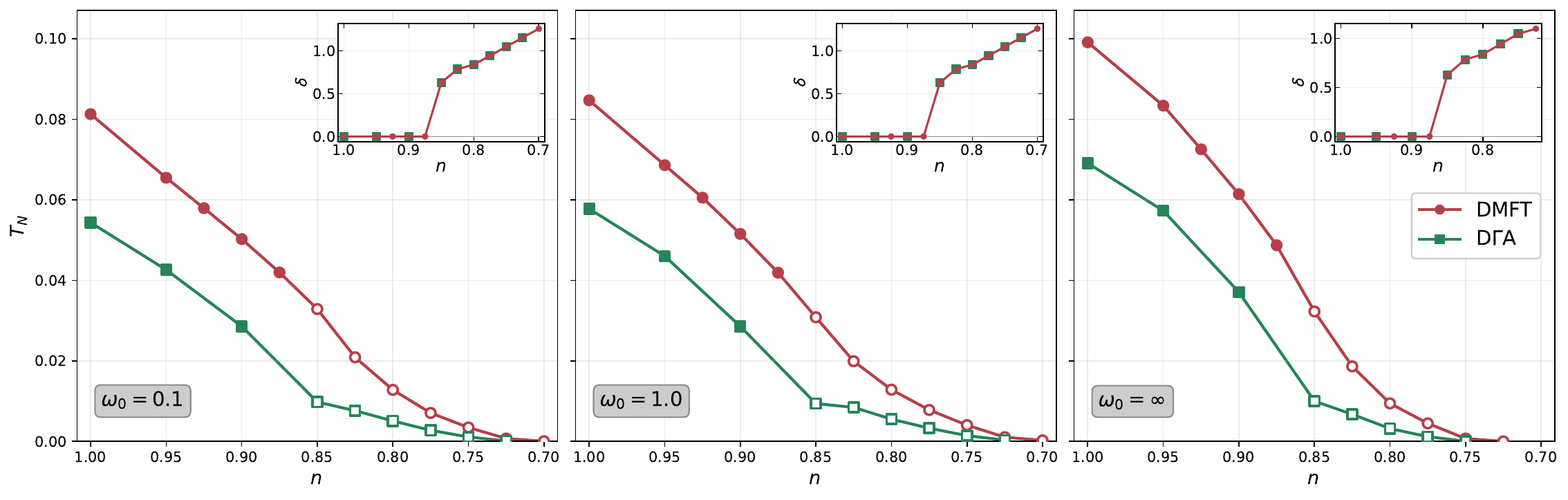}
    \caption{Doping dependence of the DMFT (red) and ladder-D$\Gamma$A (green) Néel temperature $T_N$ for $\omega_0=0.1$ (left panel), $\omega_0=1$ (middle panel) and $\omega_0=\infty$ (right panel), with $U=3$ and $\lambda=1$. Empty markers indicate incommensurate ordering vectors for the spin response, $\mathbf{Q}=(\pi,\pi,\pi-\delta)$. The incommensurability $\delta$ as a function of filling $n$ for each case is shown in the respective insets.}
    \label{fig:ohf_TN}
\end{figure*}

\begin{figure}
    \centering
    \includegraphics[width=1.0\linewidth]{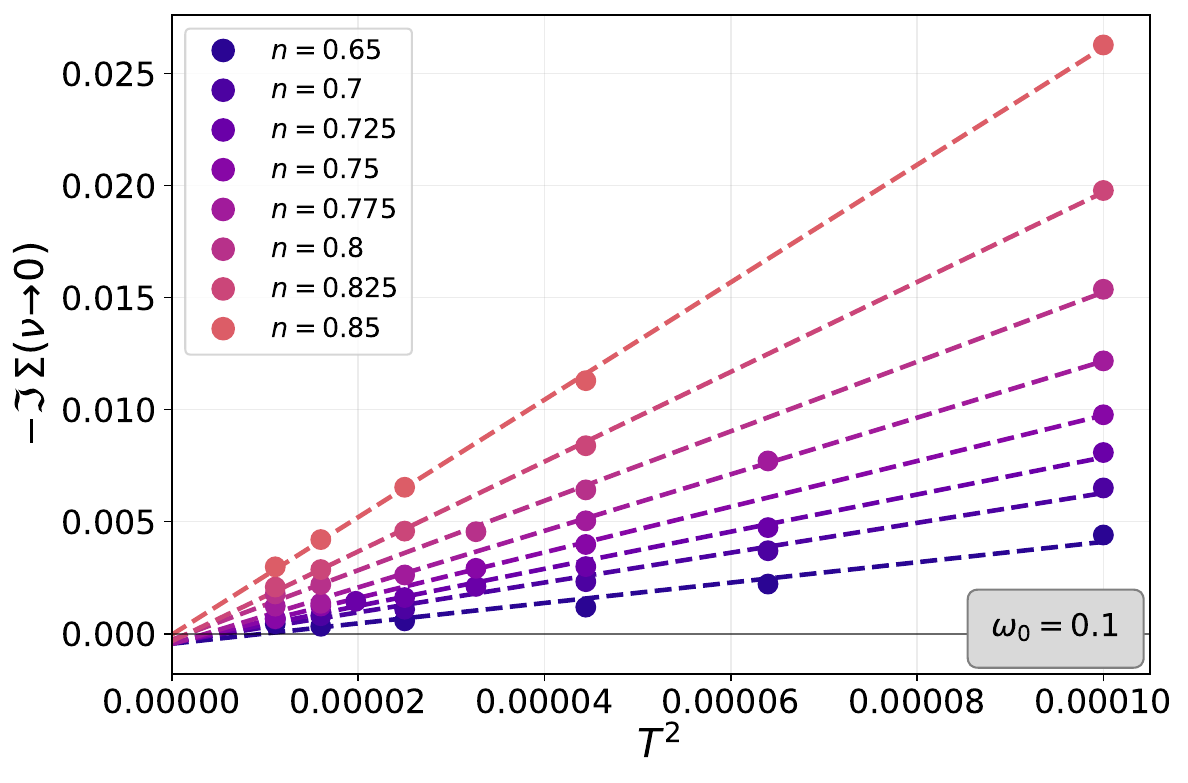}
    \caption{Temperature dependence of the DMFT scattering rate $-\Im \Sigma(\nu\to0)$ for $\omega_0=0.1$, $U=3$ and $\lambda=1$ in the quantum-critical regime. The data are extrapolated to zero temperature with a quadratic fit to the three lowest temperature points available.}
    \label{fig:gamma_T}
\end{figure}

A reduction of $T_N$ with doping is intuitively expected. Yet, the effect of doping is richer than a simple reduction and it is reflected in a drastically different temperature dependence with respect to half-filling. Close to the QCP associated with the disappearance of the antiferromagnetic state as a function of doping, it was shown that the Kohn-anomalies in the underlying Fermi-surface lead to a qualitatively different phase-diagram and critical behavior comparing RPA \cite{fusMA}, DMFT \cite{fus2026} and even $\lambda$-corrected susceptibilities \cite{schäfer2017}. These anomalies originate from so-called Kohn-points (KP), which are pairs of points on the Fermi-surface with antiparallel Fermi velocities coupled by the ordering vector of the dominant fluctuations. Indeed, as already pointed out in Refs.~\cite{schäfer2017, stepanenko2017, fus2026}, the existence of isolated or lines of KPs in 3D can trigger non-analyticities in the physical response functions in the zero temperature limit \textit{in addition} to those arising from quantum-critical fluctuations. These alter the temperature dependence of the magnetic susceptibilities $\chi_m\propto T^{-\gamma}$ and the associated correlation length $\xi\propto T^{-\nu}$ w.r.t. the expected Hertz-Millis-Moriya \cite{loehneysen2007} values ($\gamma=3/2, \, \nu=3/4$) modifying the Fisher relation for magnetic quantum phase transitions (QPT) in 3D, $\gamma=2\nu$, linking the two. Consequently, in the quantum critical regime, the existence of lines of KPs, which is relevant for the simple cubic case considered in our work, an \textit{inversion} of the Fisher relation is observed, yielding $\gamma=1/2 \, \nu$ with $\gamma=1/2$ and $\nu=1$ (which corresponds to $\chi_m(T)^{-1}\sim\sqrt{T}$ and $\xi_m(T)^{-1}\sim T$). Remarkably, this temperature dependence of the response function in the quantum-critical regime will also be reflected in a parabolic dependence of $T_N$ on doping according to $T_N\propto (n-n_c)^2$ \cite{fus2026}, with $n_c$ being the density of the QCP. This is in stark contrast with the generic square-root dependence observed in the absence of any Kohn-anomalies \cite{sachdev1999}.

Similarly to the half-filled model, we will analyze here the effect of finite el-ph coupling for the three considered phonon frequencies on the quantum-critical behavior of the magnetic susceptibilities as well as the general doping dependence of the antiferromagnetic transition temperature. In this case, however, we will fix the effective interaction to $\widetilde{U}=U-\lambda=2$ with $U=3, \lambda=1$, which corresponds to the maximum of the antiferromagnetic dome in the anti-adiabatic limit or purely electronic Hubbard model and focus on the doping dependence.

In order to address this question we find it instructive to discuss the temperature dependence of the magnetic response functions with varying densities before the doping dependence of $T_N$. To this aim, we plot the inverse static susceptibility $\chi^{-1}_m(T)$ at its ordering vector $\mathbf{Q}(T)$ calculated within DMFT (upper panel) and D$\Gamma$A (lower panel) in Fig.~\ref{fig:chi_T} for different fillings and the three characteristic frequencies, shown in the three columns. With increased doping we observe the aforementioned deviation from the typical linear or square temperature dependence in DMFT and D$\Gamma$A, respectively towards the (Kohn-line-driven) $\sim\sqrt{T}$ behavior in both cases and all phonon frequencies.

Evidently, in this case, extracting the critical temperature for the transition will require fitting the susceptibilities according to $\chi_m\propto T^{-1/2}$. We opted for this square-root fit for all densities below $n=0.825$ for both DMFT and D$\Gamma$A. Further details on the fitting procedure are provided in Appendix~\ref{app:fit}. The Néel temperature extracted this way as a function of the density is shown in Fig.~\ref{fig:ohf_TN} with the different panels distinguishing the characteristic frequencies. We denote the transition towards an incommensurate (or spin-density-wave (SDW)) state, where the ordering vector deviates from the perfect nesting case to $\mathbf{Q}=(\pi,\pi,\pi-\delta)$ with empty markers. The evolution with doping of this incommensurability $\delta$ is plotted in the corresponding insets and shows qualitatively equal behavior in all cases. 

Notably, as also already observed in the temperature dependence of the magnetic response functions, the parabolic shape of the $T-n-$phase diagram in the quantum-critical regime is not altered by the inclusion of phonons. In principle, one could expect that in the adiabatic regime of small $\omega_0$ the phonons would behave \textit{classically} below some temperature-dependent energy-scale. As discussed extensively in Refs.~\cite{freericks1993,millis96,freericks2003} the classical (adiabatic) Holstein model shares a close analogy with disordered systems, which can show a finite scattering rate even at zero temperature. If this was the case in our calculations, such a non-Fermi-liquid behavior could substantially alter the Fermi-surface properties and diminish or even fully suppress the effect of KPs. 

However, the square-root dependence of the magnetic susceptibilities (cf. Appendix~\ref{app:fit}) and the parabolic shape of the quantum-critical region indicate, even in the presence of adiabatic phonons, a low-frequency self-energy behavior incapable of destroying the sharpness and the geometry of the underlying Fermi-surface. This is demonstrated explicitly in Fig.~\ref{fig:gamma_T}, where we show the temperature dependence of the DMFT scattering rate $-\Im\Sigma(\nu\to0)$, obtained from a polynomial fit to the Matsubara self-energy, for densities close to the QCP and for the most adiabatic case, namely $\omega_0=0.1$ considered here. The scattering rate itself is fitted with a quadratic function and shows a vanishing $T=0$ value. The apparent absence of any finite scattering for zero-temperature along with the temperature dependence of the magnetic susceptibility explains why the quantum-critical behavior for all characteristic phonon frequencies remains fully controlled by the lines of KPs. We stress that in DMFT and ladder-D$\Gamma$A the mere distinction between the physical response-functions is the $\lambda$-correction scheme, i.e. in both cases the DMFT self-energy is used to dress the Green's functions in the BSE. This is in contrast to the self-consistent ladder-D$\Gamma$A  \cite{kaufmann2021} where the feedback of a non-local self-energy in the BSE is considered. In principle, even without classical phonons, including self-energy corrections in this way could potentially affect the Fermi-surface properties.

Additionally, we observe that in close analogy to the half-filled case, spatial correlations included in D$\Gamma$A lead to a remarkable reduction of the transition temperature with respect to DMFT for all doping values considered. The role of more adiabatic phonons seems to be a suppression of magnetic fluctuations, and hence of $T_N$, close to half-filling,  both for DMFT and D$\Gamma$A. For larger dopings, although smaller values of $\omega_0$ show no substantial effect on the antiferromagnetic transition temperature, a minor enhancement of $T_N$ leading to a slight shift of the QCP to smaller densities can be noted. This becomes also evident in the different scales of the inverse spin susceptibilities of Fig.~\ref{fig:ohf_TN} for different $\omega_0$ in the large doping region. There, close to the QCP, more adiabatic phonons exhibit overall smaller values for $\chi^{-1}$ or equivalently a larger magnitude for the response function. We interpret this behavior in light of Ref.~\cite{sangiovanni2006dynamics}, which showed that the effect of the el-ph coupling \(\lambda\) depends sensitively on both the ratio \(\omega_0/U\) and the filling. Near half filling, Eq.~\eqref{equ:Ueff_screening} implies that phonon-mediated screening becomes more effective with increasing \(\omega_0\) and is strongest in the antiadiabatic limit. Because the chosen interaction, $U=3$ lies on the strong-coupling side of the antiferromagnetic dome where a Heisenberg picture can be applied and  $T_N \sim 1/U$, this enhanced screening leads to an increased Néel temperature. This explains the pronounced frequency dependence of \(T_N\) at half filling. Upon doping, however, more adiabatic phonons screen the local repulsion less efficiently and are therefore less effective in restoring quasiparticle coherence. Electronic correlations and the magnetic response consequently remain stronger, allowing antiferromagnetic order to persist slightly farther from half filling and shifting the magnetic QCP. The quantitative effect is however very small.

\subsubsection{Fermionic Self-energy}

\begin{figure*}
    \centering
    \includegraphics[width=\textwidth]{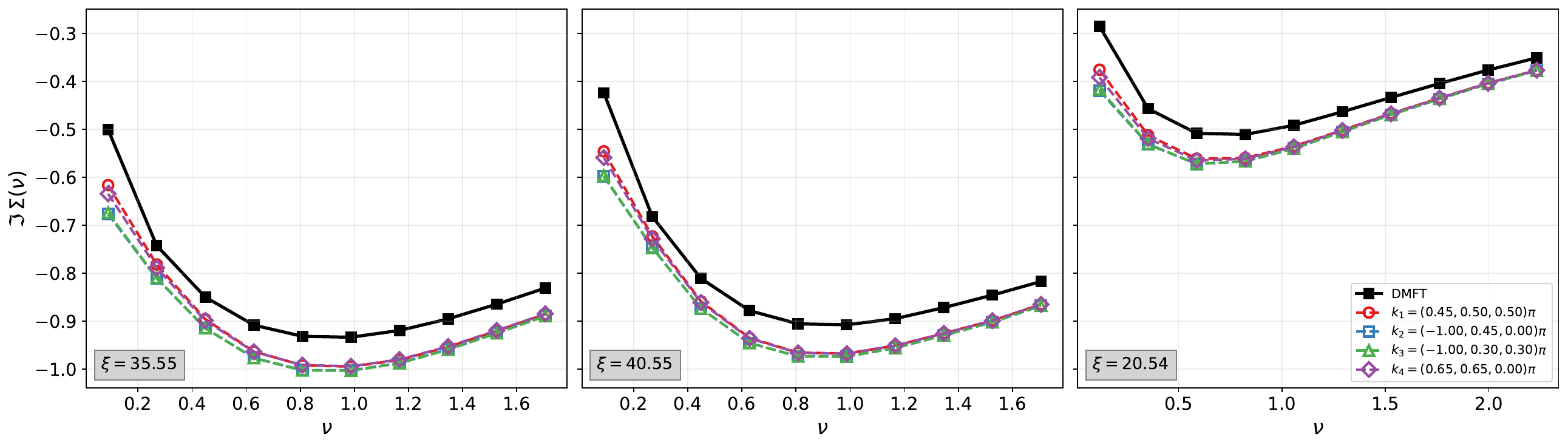}
    \caption{Comparison of the ladder-D$\Gamma$A self-energy (colored) at various momenta on the Fermi-surface with the local DMFT one (black). The data are obtained for a filling of $n=0.9$ for $\omega_0=0.1$ (left panel), $\omega_0=1$ (middle panel) and $\omega_0=\infty$ (right panel), with $U=3$ and $\lambda=1$ at the lowest available temperatures, $\beta=35$ for finite $\omega_0$ and $\beta=26.75$ for the antiadiabatic limit. The gray box shows the respective value of magnetic correlation length $\xi$ obtained from an OZ fit to the ladder-D$\Gamma$A susceptibilities.}
    \label{fig:sigma_xi}
\end{figure*}

The preceding analysis showed how the phonon frequency affects the magnetic fluctuations and phase diagram of the doped model. We now examine how these fluctuations feed back onto the electrons by comparing the D$\Gamma$A self-energy with its local DMFT counterpart in Fig.~\ref{fig:sigma_xi}. The comparison is made at $n=0.9$, $U=3$, and $\lambda=1$ for $\omega_0=0.1$, $1$, and $\infty$. The two finite-frequency results were obtained at $\beta=35$, while the antiadiabatic result was obtained at $\beta=26.75$. 

In order to put the data in the proper context, we recall that the magnetic correlation length $\xi$ measures the spatial extension of the antiferromagnetic correlations. Hence, it provides a measure of the proximity to the magnetic phase transition: the larger $\xi$, the closer the system is to the instability. The set of parameters for the three phonon frequencies was chosen such that the correlation lengths (indicated in the grey boxes of Fig.~\ref{fig:sigma_xi}) share the same order of magnitude and are relatively large or closer to the antiferromagnetic transition. From Fig.~\ref{fig:sigma_xi}, displaying the imaginary part of the electron self-energy, we immediately observe that more adiabatic phonons produce stronger low-frequency scattering already at the DMFT level consistent with Ref.~\cite{sangiovanni2006dynamics}, while D$\Gamma$A yields a comparable enhancement of the scattering rate ($\Im \Sigma(\nu\to0,\mathbf{k})$) in all three datasets. At the same time, the self-energies at the chosen Fermi-surface momenta remain very similar, indicating no pronounced momentum differentiation for the parameters considered here. This feature appears qualitatively consistent with previous results obtained in 3D for the Hubbard model \cite{toschi2007,rohringer2016impact}.

\subsection{Superconductivity in 2D}\label{sec:sc}
As a second illustrative application of dynamical-$U$ ladder D$\Gamma$A, we study superconductivity in the two-dimensional Hubbard-Holstein model. This allows us to examine how the phonon-induced, frequency-dependent interaction modifies an electronically driven $d$-wave instability and its transition temperature.

Previous dynamical cluster approximation (DCA) studies of the Hubbard--Holstein model indicate that phonons can enhance the pairing vertex, while the concomitant increase in quasiparticle scattering generally suppresses $T_c$ \cite{macridin2006,khatami2008,macridin2009,macridin2012}. The occurrence of a similar trend in the proximity of the Mott metal-to-insulator transition (MIT) has been also recently discussed in Ref.~\cite{moghadas2026}. Unconventional superconductivity in the doped two-dimensional Hubbard model has been studied extensively as a paradigmatic example of pairing driven by electronic correlations. A reliable description of this problem requires a simultaneous treatment of strong local dynamical correlations and nonlocal spatial fluctuations, as provided by cluster methods and diagrammatic extensions of DMFT \cite{rohringer2018diagrammatic}. These approaches have identified antiferromagnetic fluctuations as a central ingredient of $d$-wave pairing \cite{lichtenstein2000,kyung2003,maier2005,capone2006competition,haule2007,hafermann2009,sordi2012,gull2013,otsuki2014,gunnarson2015,kitatani2015,gunnarson2016,vucicevic2017,kitatani2019,kitatani2022}. Within D$\Gamma$A, in particular, retaining the full frequency dependence of the irreducible pairing vertex turned out to be essential to obtain realistic estimates of $T_c$, whereas simpler approximations such as fluctuation exchange (FLEX) tend to overestimate the transition temperature \cite{kitatani2015,kitatani2019}. 

Here, we revisit the analysis of the competing microscopic processes in the Hubbard-Holstein model, namely pairing vs. pair-breaking, using ladder D$\Gamma$A, which operates directly in the thermodynamic limit and can describe spatial correlations beyond a finite cluster size. In particular, we investigate how the strength and characteristic frequency of the el-ph interaction affect the pairing vertex, quasiparticle coherence, and the resulting $d$-wave transition temperature.

\subsubsection{Superconductivity in ladder-D$\Gamma$A}

The most general approach to SC is a formulation in Nambu space which allows us to access also the symmetry-broken state via the inclusion of the so-called anomalous Green's function, which can become extremely heavy, especially in extensions of DMFT such as D$\Gamma$A.

However, a linearization of the anomalous components allows one to compute the critical temperature $T_c$ and some properties of the emerging superconducting gap function close to $T_c$ based purely on normal state quantities.  In order to study SC within the ladder-D$\Gamma$A method we adopt here the formulation introduced in Ref.~\cite{kitatani2019} for the pure Hubbard model, which avoids the computationally expensive treatment of the parquet equations as well as a diagrammatic formulation based on DMFT in the Nambu basis. The transition temperature can be estimated by solving the linearized gap equation, an eigenvalue problem, for the pairing channel \cite{otsuki2014, kitatani2015}
\begin{equation}\label{equ:eliashberg}
    \epsilon_{SC} \, \Delta(k) = -\frac{1}{\beta}\sum_{k'}\Gamma_{pp}(k,k',q=0)G(k')G(-k')\Delta(k')
\end{equation}
where $\Gamma_{pp}$ denotes the irreducible vertex in the particle-particle (\textit{pp}) channel, $\Delta(k)$ can be seen as the anomalous Green's function, and $\epsilon_{SC}$ is the superconducting eigenvalue, with $\epsilon_{SC} \to 1$ signaling a superconducting instability. As detailed in Refs~\cite{kitatani2019,kitatani2022,jacob2026}, the pairing vertex of ladder-D$\Gamma$A is built of \textit{ph}-ladders, which mediate non-local spin fluctuations thereby acting as the \textit{pairing glue}. More specifically using the ladder-D$\Gamma$A full vertices, $\widetilde{F}$, of Eq.\eqref{eq:F_upup} and \eqref{eq:F_updn}, which were also the main components of SDE in Eq.~\eqref{equ:sde_dga}, we can construct the $\Gamma_{pp}$ in the singlet or triplet channels, $\alpha \in \{s/t\}$, by expressing the \textit{ph}-frequency arguments accordingly in the \textit{pp}-convention \cite{rohringer2014phd}
\begin{equation}\label{equ:gamma_pp}
    \begin{aligned}
        &{}  \Gamma^{\alpha}_{pp}(k,k',q=0) \equiv \Gamma^{\alpha}_{pp, \mathbf{Q=k'-k}}(\nu,\nu') \\
        &{}= \frac{1}{2} F_{d,\mathbf{k'-k}}^{\nu,-\nu,\nu'-\nu} - \left( \frac{3}{2} + \delta_{\alpha,t}  \right) F_{m,\mathbf{k'-k}}^{\nu,-\nu,\nu'-\nu}  \\
        &{} - F_{loc,\uparrow\downarrow}^{\nu,-\nu,\nu'-\nu} - \Phi_{loc,pp,\uparrow\downarrow}^{\nu,\nu'\omega=0} + (1 - 2\delta_{\alpha,t}) (k' \to -k')
    \end{aligned}
\end{equation} 
Here, the last term of the last line denotes symmetrization (antisymmetrization) w.r.t.~the second fermionic four-momentum $k'$. Throughout the remainder of this section we will be interested in uniform ($\mathbf{q}=0$) even-frequency, $d$-wave pairing in the singlet channel. Pair-density waves or other exotic (e.g., odd-frequency, triplet, d-wave) SC are beyond the scope of this work.

\subsubsection{d-wave SC in the Hubbard-Holstein model}
\begin{figure}[h!]
    \centering
    \includegraphics[width=1.0\linewidth]{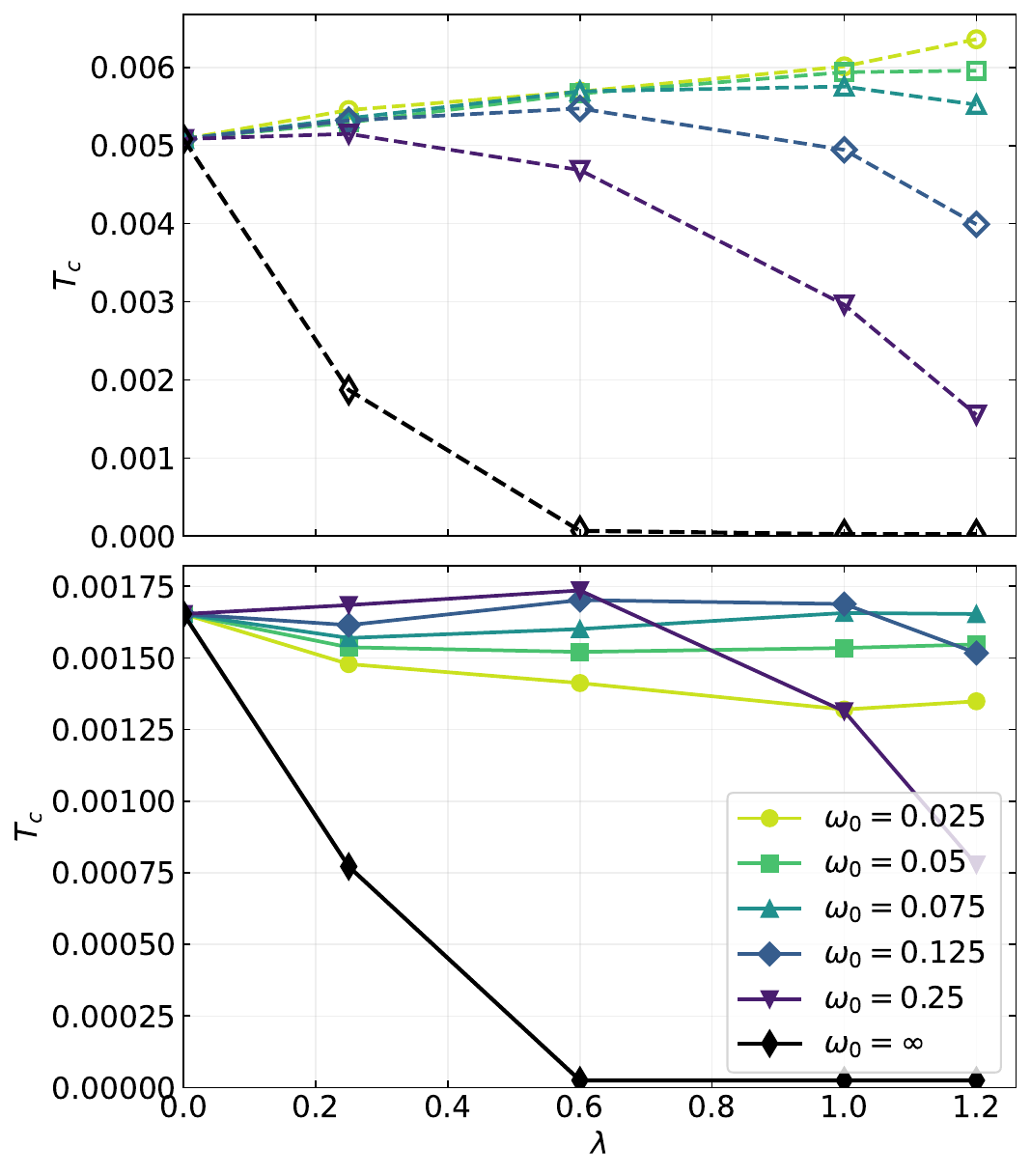}
    \caption{$\lambda$-dependence of the $d$-wave superconducting transition temperature $T_c$ for the 2D HHM for various phonon frequencies $\omega_0$ with DMFT Green's function (upper panel) and with D$\Gamma$A self-energy feedback (lower panel). }
    \label{fig:TC}
\end{figure}

\begin{figure}[h!]
    \centering
    \includegraphics[width=1.0\linewidth]{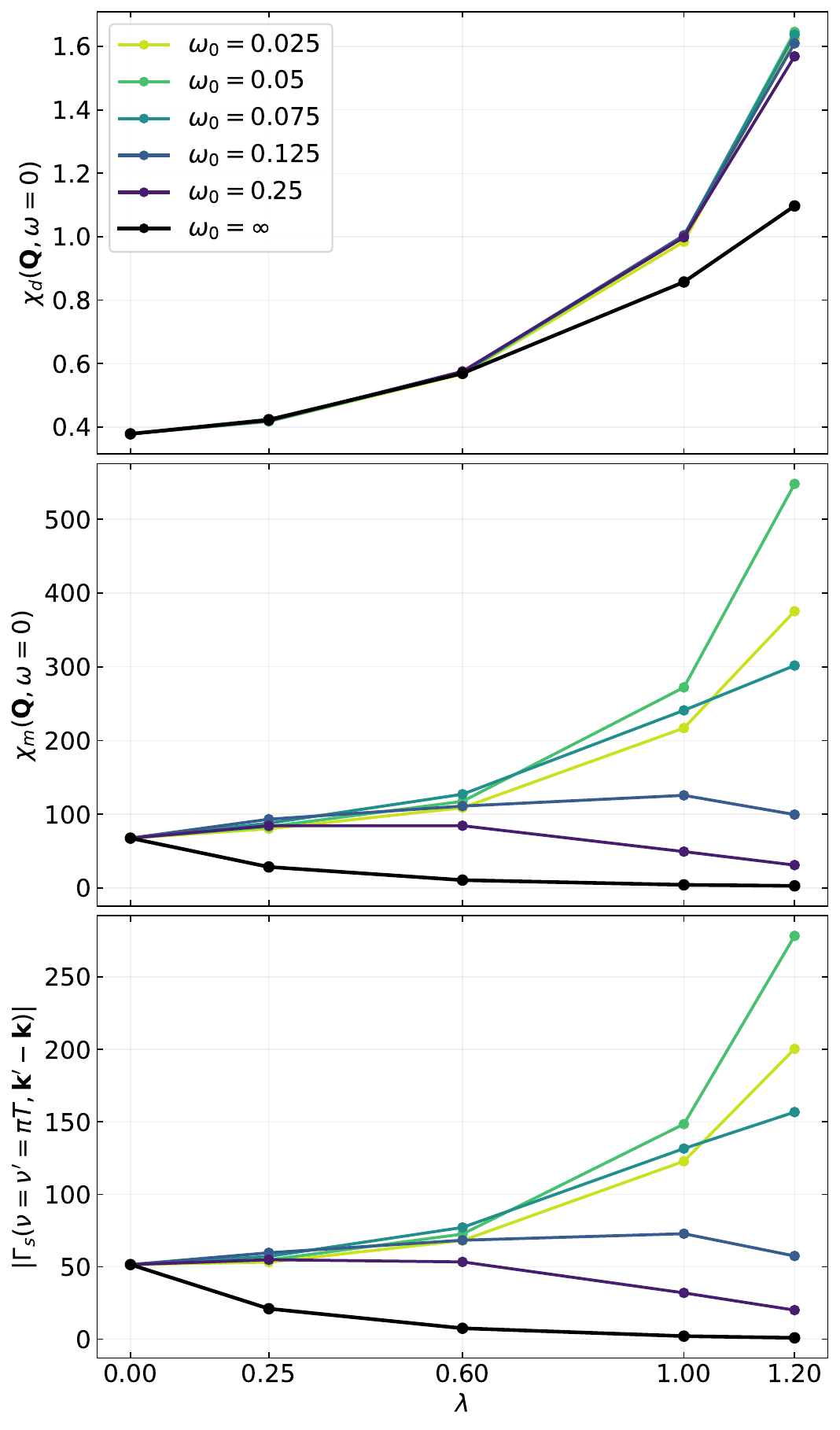}
    \caption{Comparison of the $\lambda$-dependence of the D$\Gamma$A susceptibilities in the density (upper panel) and magnetic (middle panel) with the irreducible vertex in the singlet channel (lower panel) evaluated at the lowest fermionic frequency $\nu=\pi T$ and the ordering vector of the spin response function, i.e., $\mathbf{k'-k}=\mathbf{Q}$.}
    \label{fig:chi_r_gamma}
\end{figure}

\begin{figure*}
    \centering
    \includegraphics[width=1.0\linewidth]{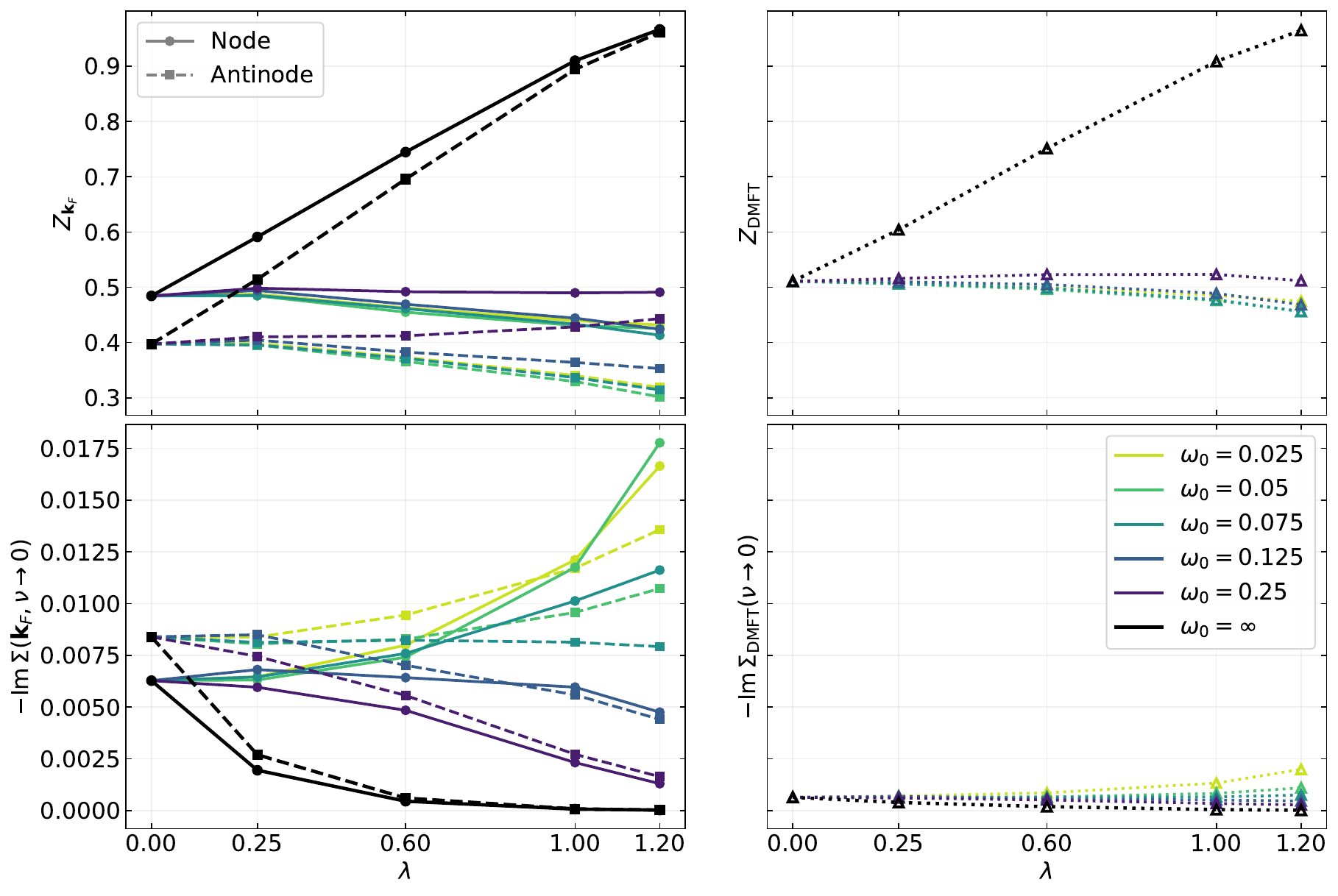}
    \caption{Comparison of D$\Gamma$A (left column) and DMFT (right column) quasiparticle weight $Z$ (upper row) and scattering rate $\Im \Sigma(\nu\to0)$ at $\beta=100$ and for various phonon frequencies. In the left column the nodal region is marked with bold lines, whereas the antinodal region is denoted by dashed lines.}
    \label{fig:QP_params}
\end{figure*}
    
We now specialize to $d$-wave superconductivity in the Hubbard--Holstein model on the square lattice. To allow a direct comparison with Ref.~\cite{kitatani2019}, we consider a nearest-neighbor dispersion with $t'=t''=0$, a Hubbard interaction $U=1.5$, and a filling $n=0.825$ close to optimal doping. Keeping these electronic parameters fixed, we investigate how the superconducting transition temperature evolves with the el-ph coupling strength $\lambda$ and characteristic frequency $\omega_0$. The additional phonon-mediated attraction can enhance nonlocal charge fluctuations, which enter the pairing interaction through the particle--hole ladders and may therefore influence the $d$-wave instability. To avoid artificially suppressing these fluctuations (see discussion in Sec.~\ref{sec:lamcorr}), we apply the Moriya correction only in the magnetic channel throughout this section. This choice allows us to fully retain the phonon-induced charge response in our investigation and assess its contribution to the superconducting pairing interaction. An example of the leading Eliashberg eigenvalues and eigenvectors along with their frequency and momentum dependence is provided in Appendix~\ref{app:bcs_comparison}. There we also compare the characteristic energy scales of the D$\Gamma$A superconducting gap function with those of a conventional $d$-wave BCS model.

By solving the linearized Eliashberg equation for several temperatures and phonon frequencies, we determine the dependence of the $d$-wave transition temperature on the el-ph coupling $\lambda$. Figure~\ref{fig:TC} shows the resulting $T_c$ for different characteristic phonon frequencies $\omega_0$. In the upper panel, the Eliashberg equation is evaluated using the momentum-dependent DMFT Green's functions thereby neglecting the feedback of the non-local D$\Gamma$A self-energy and allowing us to isolate the effect of the pairing vertex. In the lower panel, this feedback is consistently included and the corresponding Green's functions of the ladder-D$\Gamma$A are used instead.

The main results which we observe are the following: without self-energy feedback, sufficiently adiabatic phonons enhance $T_c$ as $\lambda$ increases, whereas faster phonons progressively suppress the superconducting instability upon approaching the antiadiabatic limit. Once the D$\Gamma$A self-energy is included, however, the enhancement is largely compensated, and the el-ph interaction displays a moderate enhancement only for some frequencies while the general effect is a suppression of $T_c$. Moreover, while a larger $\omega_0$ initially favors superconductivity at weak el-ph coupling, this trend seems to eventually reverse at the strongest $\lambda$-values considered here, namely at $\lambda=1.2$. There the highest transition temperatures occur at more adiabatic phonon frequencies.

Further insight into this behavior can be obtained by examining the two main components of the Eliashberg equation: the pairing vertex $\Gamma_{pp}$ and the \textit{pp}-propagator $G(k)G(-k)$ containing the D$\Gamma$A self-energy. In Fig.~\ref{fig:chi_r_gamma}, we compare the density and magnetic response functions, $\chi_d$ and $\chi_m$, evaluated at their dominant transfer momentum $\mathbf{Q}$, with the magnitude of the corresponding singlet pairing vertex evaluated at the same momentum in the \textit{pp}-channel $\mathbf{Q}\equiv\mathbf{k'-k}$. We immediately observe that the evolution of $\Gamma_{pp}$ closely follows that of the magnetic susceptibility, particularly at larger $\lambda$, where both quantities develop a pronounced dependence on $\omega_0$. As in the three-dimensional antiferromagnetic case, faster phonons screen the Hubbard interaction more efficiently, thereby suppressing magnetic fluctuations and weakening the pairing interaction. More adiabatic phonons, by contrast, preserve or even enhance the magnetic response and consequently increase the pairing amplitude. This behavior explains the enhancement of $T_c$ found when self-energy feedback is neglected. Although the density response also grows substantially with $\lambda$, considering the respective scales of the two responses reveals that it remains much smaller than the magnetic susceptibility and provides only a subleading contribution to pairing in the parameter regime considered here. Its influence may however become more important closer to a charge-ordering instability, which is not generally excluded by the Mermin-Wagner theorem in two dimensions.

From the above discussion it becomes clear that the suppression of $T_c$ in the full D$\Gamma$A calculation must originate primarily from the self-energy contribution to the \textit{pp}-bubble term $G(k)G(-k)$. Figure~\ref{fig:QP_params} shows the quasiparticle weight $Z$ and the zero-frequency scattering rate $-\Im\Sigma(\nu\to0)$ as functions of $\lambda$ and $\omega_0$. The D$\Gamma$A results in the left column distinguish between nodal (bold lines) and antinodal (dashed lines) momenta, while the corresponding local DMFT quantities are shown in the right column. Only a moderate momentum-dependence is observed, consistent with the filling $n=0.825$ lying outside the strongly momentum-differentiated pseudogap regime closer to half-filling. In this regime, we observe that the overall dependence on $\lambda$ and $\omega_0$ is qualitatively similar in DMFT and D$\Gamma$A. Specifically, in the antiadiabatic limit, the phonon-mediated interaction primarily screens the Hubbard repulsion, causing $Z$ to increase and the scattering rate to decrease with increasing $\lambda$. For slower phonons, this trend is gradually reversed: static screening becomes less effective, while the conventional phonon-induced mass renormalization reduces $Z$ and enhances quasiparticle scattering. Consistent with Ref.~\cite{sangiovanni2006dynamics}, the nonmonotonic $\omega_0$-dependence indicates the presence of an intermediate low-frequency regime in which the quasiparticle renormalization is strongest. The similar trends observed in DMFT and D$\Gamma$A hold, however, only at the qualitative level. Quantitatively, the D$\Gamma$A scattering rate is almost an order of magnitude larger than its DMFT counterpart, highlighting the substantial amount of additional scattering generated by nonlocal fluctuations.

Taken together, these results reveal a competition between pairing and decoherence. Adiabatic phonons strengthen the magnetic contribution to the pairing vertex, but simultaneously reduce quasiparticle coherence and, most importantly, enhance scattering. Faster phonons improve coherence through screening, while weakening the magnetic pairing interaction. The resulting balance of pairing versus pair-breaking, consistent with the DCA results of Ref.~\cite{macridin2006,macridin2009,macridin2012}, explains why the full D$\Gamma$A transition temperature (including the non-local self-energy feedback) is only weakly modified w.r.t. the pure Hubbard model or rather generally suppressed by adding the el-ph coupling.

\section{Conclusion}\label{sec:sum}
In this work we extended the equations and algorithm of the dynamical vertex approximation (D$\Gamma$A) in its ladder variant to systems with on-site dynamical interactions, allowing a common diagrammatic framework to describe retarded interactions and nonlocal electronic correlations.   

The requirements for the presented algorithm, termed dynamical-$U$ ladder-D$\Gamma$A, are rooted in principle on two fundamental extensions w.r.t.~the existing scheme, namely (i) an appropriate formulation of the ladder diagrams incorporated in the Schwinger-Dyson equation of motion (SDE) and the (ii) Moriya-$\lambda$ correction of its main constituents, namely the DMFT lattice response functions. This correction has to be done for both the magnetic and density channels in order to assure the correct high-frequency asymptotics of the resulting D$\Gamma$A self-energy. Hence, the formulation of the SDE in terms of these corrected susceptibilities results in a nontrivial dependence on fermionic and bosonic frequencies, which, as a consequence, deviates significantly from its definition in the absence of dynamical interactions. 

To illustrate the applicability of the proposed method, we exploited it for studying one of the simplest models which can exhibit a frequency-dependent effective electronic interaction, namely the Hubbard-Holstein model. In particular, we investigated, in ladder-D$\Gamma$A, the effect of phonons for finite characteristic frequencies on the antiferromagnetic (AFM) transition in 3D as well as the onset of $d$-wave superconductivity (SC) in 2D, and used the antiadiabatic limit of the model, where it maps to a purely electronic Hubbard model, to compare against existing results.

In 3D, consistent with previous literature, we observed that non-local correlations generically reduce the Néel temperature ($T_N$) w.r.t.~DMFT, at and out of half-filling \cite{rohringer2018diagrammatic}. Spatial fluctuations do not, however, in ladder-D$\Gamma$A, qualitatively change the role of phonons on the AFM transition. In agreement with previous DMFT literature, for the presently considered parameter-regime, we find that the main role of the phonons is twofold: A screening of the Hubbard-interaction $U$,  whose strength depends on the ratio of $\omega_0/U$, and the opening of an additional scattering channel, which further contributes to the decoherence of the quasiparticles (QP) \cite{sangiovanni2005, sangiovanni2006dynamics}. At half-filling, for effectively repulsive interactions, the former mechanism prevails. There, we observed that adiabatic phonons are less efficient in opposing the repulsive interaction leading to a rather weak dependence of $T_N$ on the electron-phonon (el-ph) coupling strength $\lambda$, compared to the antiadiabatic limit, where the el-ph coupling directly competes with $U$. Doping the system eventually renders the effect of $U$ on the system less relevant. Therefore, we recover the generically expected role of the phonons, namely the localization and decoherence of the QP, which can lead, in turn, to longer-lived spin-fluctuations. Consequently, close to half-filling, where the phonons oppose $U$, the AFM transition is suppressed. In contrast, further away from half-filling, more adiabatic phonons induced a stronger magnetic response and, hence, larger values of $T_N$, shifting the associated quantum-critical-point (QCP) to slightly larger dopings. 

The qualitative behavior in the proximity of the QCP, which for the lattice dispersion considered is dictated by Kohn-points (KP) physics in the pure Hubbard model on 3D lattice \cite{fusMA,fus2026,schäfer2017,stepanenko2017}, showed no particular dependence on the phonon-frequency. In principle, purely classical oscillators could yield a finite quasi-particle (QP) scattering even at $T=0$ \cite{millis96, freericks1993, freericks2003} and, hence, potentially destroy the KP-driven Fermi-surface geometry. Even for the most adiabatic regime of $\omega_0=0.1$ considered in this work, however, our analysis of the DMFT self-energy revealed a distinct Fermi-liquid behavior, which rules out any possible zero-temperature scattering. Together with our observation of the square-root temperature-dependence of the magnetic susceptibility and the consequent parabolic form of the $T_N(n)$, this establishes the KP-driven quantum-critical regime to be a robust feature against quantum phonons. It remains an open question left for future work, below which temperature scale, if that exists, adiabatic phonons may effectively compete against the Kohn-anomalies-driven quantum-criticality and which comparisons can be drawn to disordered systems modeled for instance with cluster-extensions of DMFT \cite{tscheppe2025}. 

As a second application of the method, we studied the dependence of the 2D $d$-wave superconducting transition temperature on the el-ph coupling strength $\lambda$ as well as the characteristic frequency $\omega_0$. The D$\Gamma$A formulation of the linearized Eliashberg kernel employed to obtain $T_c$ revealed a trend qualitatively consistent with previous literature on the superconductivity in the Hubbard-Holstein model \cite{macridin2006, khatami2008, macridin2009}. More specifically, here, we have investigated how the behavior of the critical temperature is governed by the competition between a pair-forming and pair-breaking effect mediated by phonons. On one hand, more adiabatic phonons tend to significantly enhance nonlocal magnetic fluctuations, resulting in a substantially larger pairing amplitude upon increasing $\lambda$, while simultaneously increasing the QP scattering. On the other hand, faster phonons improve the coherence of the QP by an efficient screening of the Hubbard interaction. This happens, however, at the cost of a drastically suppressed pairing-interaction, due to weaker nonlocal magnetic fluctuations. Such a competition explains the weak and non-monotonic dependence of the D$\Gamma$A transition temperature on the phononic properties. Future studies should investigate the role of strong density fluctuations, found for instance in the proximity to phonon-driven charge order \cite{grilli1991, grilli1994, ciuchi1999, capone2004, prokofev2026, chubukov2026}.

These open questions motivate several extensions of the presented formalism. For instance, as we discussed, one limitation is represented by the Moriya-$\lambda$ correction in the density channel. In two dimensions, imposing the corresponding local two-particle consistency condition suppresses any finite-temperature divergence of the charge susceptibility in a similar fashion as it would happen in TPSC \cite{dare1996,vilk1997,delre2019,titvinidze2025,leusch2026}, although charge ordering and phase-separation are not generally prohibited by the Mermin-Wagner theorem. The present approach may therefore underestimate charge fluctuations and their contribution to superconducting pairing, particularly near charge-ordering or phase-separation instabilities. This feature is especially relevant close to a Mott transition, where single-site and cluster DMFT studies have found enhanced charge fluctuations that can strongly renormalize the effective el-ph coupling and potentially influence superconductivity in multiorbital systems such as iron-based superconductors \cite{arrigoni2003, demedici2017hund, reitner2020attractive, meixner2026, moghadas2026}. 

One promising route to be undertaken is, then,  self-consistent D$\Gamma$A (sc-D$\Gamma$A) \cite{kaufmann2021}, in which the nonlocal self-energy is fed back into the propagators and response functions. Indeed, by avoiding the need for an explicit Moriya-$\lambda$ correction, sc-D$\Gamma$A appears suited to describe physically allowed charge instabilities \cite{al-eryani2026diagrammatic}, while also providing a more consistent treatment of the antiadiabatic limit. Moreover, sc-D$\Gamma$A represents the most natural route  for extending the dynamical-$U$ D$\Gamma$A formalism both to multiorbital models \cite{toschi2011,galler2017}, e.g.~for realistic materials with frequency-dependent screening, as well as to systems with broken SU(2)-symmetry \cite{delre2021}, since no established Moriya-correction scheme currently exists for these cases.

These methodological enhancements would enable applications to a wider range of correlated materials in which electronic correlations, bosonic excitations, and dynamical screening occur on comparable energy scales. Overall, our results establish dynamical-$U$ ladder D$\Gamma$A as a practical framework for correlated systems with retarded interactions and demonstrate, through two illustrative examples of its application, how phonons can reshape thermodynamic instabilities through a nontrivial competition between interaction screening, nonlocal fluctuations, and quasiparticle decoherence.

\begin{acknowledgments}
We thank Aiman Al-Eryani, Sergio Caprara, Sergio Ciuchi, Dominik Fus, Eric Jacob, Thomas Schäfer, Giorgio Sangiovanni and Carlo Tortora for valuable discussions. E.M. and A.T. acknowledge support from the Austrian Science Fund (FWF), Grant 10.55776/KIN2563725 (Project P1 of the research unit QUAST, for5249, of the German Research Foundation, DFG) and Grant 10.55776/I5487 (VerDyn). S.G. acknowledges support from the Ermenegildo Zegna Founder’s Scholarship. M.C. acknowledges support of MUR via  the European Union - NextGenerationEU through PRIN 2022 (Prot. 20228YCYY7), MUR PNRR Project No. PE0000023-NQSTI, and No. CN00000013-ICSC. At this point we also want to thank the WWAQ-organization for their hospitality. Calculations have been performed on the Austrian Scientific Cluster (ASC). The authors acknowledge the use of large language models (GPT-6 Astra) to assist with English grammar and language editing, \LaTeX{} typesetting, and data postprocessing, including automated data extractions, analysis and plotting. The Flatiron Institute is a division of the Simons Foundation.
\end{acknowledgments}

\section{Appendix}

\subsection{Alternative Moriya-$\lambda$ correction scheme}\label{app:lambda_correction}

An alternative prescription for determining the $\lambda$ parameters follows from
requiring consistency between the potential energy evaluated at the one- and
two-particle levels \cite{stobbe2022}. At the one-particle level, the
potential energy per lattice site is obtained from the Migdal-Galitskii expression \cite{galitskii1958}
\begin{equation}\label{equ:epot_one_particle}
    E_{\mathrm{pot}}^{(1)}
    =\frac{1}{\beta}\sum_k
    G(k)\Sigma(k).
\end{equation}
At the two-particle level, the same quantity can be extracted from the expectation
value of the interaction part of the Hamiltonian \cite{galitskii1958}, which generically reads
\begin{equation}\label{equ:epot_two_particle_def}
    E_{\mathrm{pot}}^{(2)}
    =U\langle n_{i\uparrow}n_{i\downarrow}\rangle
    +\sum_\alpha g_\alpha
    \left\langle n_i
    (b_{i\alpha}^{\dagger}+b_{i\alpha})\right\rangle.
\end{equation}
This expression can be straight-forwardly reformulated in terms of the physical susceptibilities of the system \cite{krien2018}
\begin{align}\label{equ:epot_two_particle}
    E_{\mathrm{pot}}^{(2)}={}&
    U\left[
        \frac{n^2}{4}
        +\frac{1}{2\beta}\sum_q
        \left(
            \chi_{d}^{q}
            -\chi_{m}^{q}
            \right)
            \right]\nonumber\\
            &+\frac{1}{\beta}\sum_q V(\omega) \chi_{d}^{q} - 2V(0)\frac{n^2}{4},
\end{align}
where we can immediately identify the first line with the double occupancy term and the second one with the bosonic expectation value in Eq.~\eqref{equ:epot_two_particle_def}.

From here on out, the $\lambda$-correction for the potential energy consistency is equivalent to the one devised in \cite{stobbe2022}. The $\lambda$-corrected susceptibilities which are used to compute the potential energy in Eq.~\eqref{equ:epot_two_particle} are also implicitly in the Migdal-Galitskii expression of Eq.~\eqref{equ:epot_one_particle}. Requiring the equivalence of these two conditions, i.e., $E_{\mathrm{pot}}^{(1)}=E_{\mathrm{pot}}^{(2)}$, together with the Pauli sum-rule in Eq.~\eqref{equ:sum-rule_pauli} of the main text provides two coupled root-finding equations for $\lambda_d$ and $\lambda_m$.
For $V(\omega)=0$, Eq.~\eqref{equ:epot_two_particle} reduces to the standard
potential-energy consistent ladder-D$\Gamma$A prescription of
Ref.~\cite{stobbe2022}.

\subsection{Computational details}\label{app:comp_details}

For the numerical calculations presented in this work, we explicitly consider the single-band Hubbard-Holstein model on a hypercubic lattice with $d\in\{2,3\}$ dimensions. The corresponding Hamiltonian is given by Eq.~\eqref{equ:hh_hamiltonian} in the main text.

We employ the ladder-D$\Gamma$A algorithm for the treatment of the above Hamiltonian, which requires the local two-particle DMFT vertices as an input. These in turn can be extracted from a self-consistently determined Anderson impurity model \cite{georges1996}. The data throughout the main text and this supplementary material were obtained from a CT-QMC impurity solver in the hybridization expansion, more specifically in the segment formulation as implemented in the \textit{w2dynamics} \cite{w2dyn} code-package and \textit{ctseg} solver \cite{ctseg-triqs} as implemented in the \textit{TRIQS} \cite{triqs} software-ecosystem. 

For all QMC measurements the tight binding dispersion for the hypercubic lattice was discretized on a $48^d$ momentum-grid. After the convergence of the DMFT algorithm the one-particle quantities were recomputed with a single-shot high-statistics run with $\mathcal{O}(10^6-10^7)$ measurements. This was used as starting input for the measurement of the two-particle quantities using $\mathcal{O}(10^5-10^6)$ measurements. 

For the study of the antiferromagnetic transition in Sec.~\ref{sec:afm}, where the central quantity of interest was the susceptibility, we found a frequency grid with $n_{inner}=100$ for temperatures down to $\beta<125$, $n_{inner}=150$ for $125<\beta<200$ and $n_{inner}=200$ for $200<\beta$ with $n_{outer}=1500$ for all values of $\beta$ to be necessary for converged results. For the D$\Gamma$A momentum grids of $40^3$ for $\beta<125$ and $60^3$ for $\beta>150$ were chosen. 

In contrast, for the $d-wave$ superconductivity study of Sec.~\ref{sec:sc}, due to the relatively fast convergence of the Eliashberg eigenvalues \cite{kitatani2022}, we were able to obtain converged results with a rather small frequency grid of $n_{inner}=50$ positive frequencies with an outer asymptotic frequency box of $n_{outer}=500$ (cf. Appendix \ref{app:asymp}). For internal momentum integrals in the momentum dependent BSE, required for the lattice response functions of the ladder-D$\Gamma$A, we opted for a $120\times120$ momentum grid. 

A practical test to determine the convergence of the ladder-D$\Gamma$A algorithm w.r.t. Matsubara frequencies is comparing a local version of the SDE, Eq.~\eqref{equ:sde_dga}, with the impurity self-energy directly measured in QMC. Throughout all computations in this paper we made sure that the relative error measure in terms of the Frobenius norm remained around $\mathcal{O}(10^{-2})$ and below. Lower values for the relative error were typically found for higher temperatures, where evidently smaller frequency grids are sufficient for fully converged results.

Finally, we want to point out one caveat for our algorithm, namely the treatment of the antiadiabatic limit. As already pointed out in Sec.~\ref{sec:lamcorr}, due to the non-commutativity of (i) defining the interaction-irreducible bare interactions and (ii) of taking the antiadiabatic limit $\omega_0\to \infty$, the ladder-D$\Gamma$A algorithm with the $\lambda$-corrections scheme will not be able to continuously connect the finite $\omega_0$ regime and the antiadiabatic limit of the Hubbard-Holstein model. This is a consequence of the $\lambda$-corrections requiring a reformulation of the full vertices $F$ in terms of the physical response functions and three-leg vertices of the lattice. More specifically, we note that this reformulation is reminiscent of the single-boson-exchange formalism \cite{krien2019}, which differently from the parquet-formalism classifies diagrams in terms of irreducibility w.r.t. cutting interaction lines. The bare-interactions for the Hubbard-model or the antiadiabatic Hubbard-Holstein model are given by 
\begin{equation}\label{equ:bare_int_winf}
    \mathcal{U}_{d/m} = \pm (U + V(0))    
\end{equation}
However, we have defined the longitudinal interactions, which are the basis of the interaction-irreducibility for finite phonon frequencies $\omega_0<\infty$, as $\mathcal{U}_d = U + 2V(\omega)$ and $\mathcal{U}_m = - U$. Evidently taking the limit $\omega_0 \to \infty$ for these expressions will not yield Eq.~\eqref{equ:bare_int_winf}. Consequently, considering larger characteristic frequencies will introduce a larger error in the ladder-D$\Gamma$A formalism.

\begin{figure}[h!]
    \centering
    \includegraphics[width=1.0\linewidth]{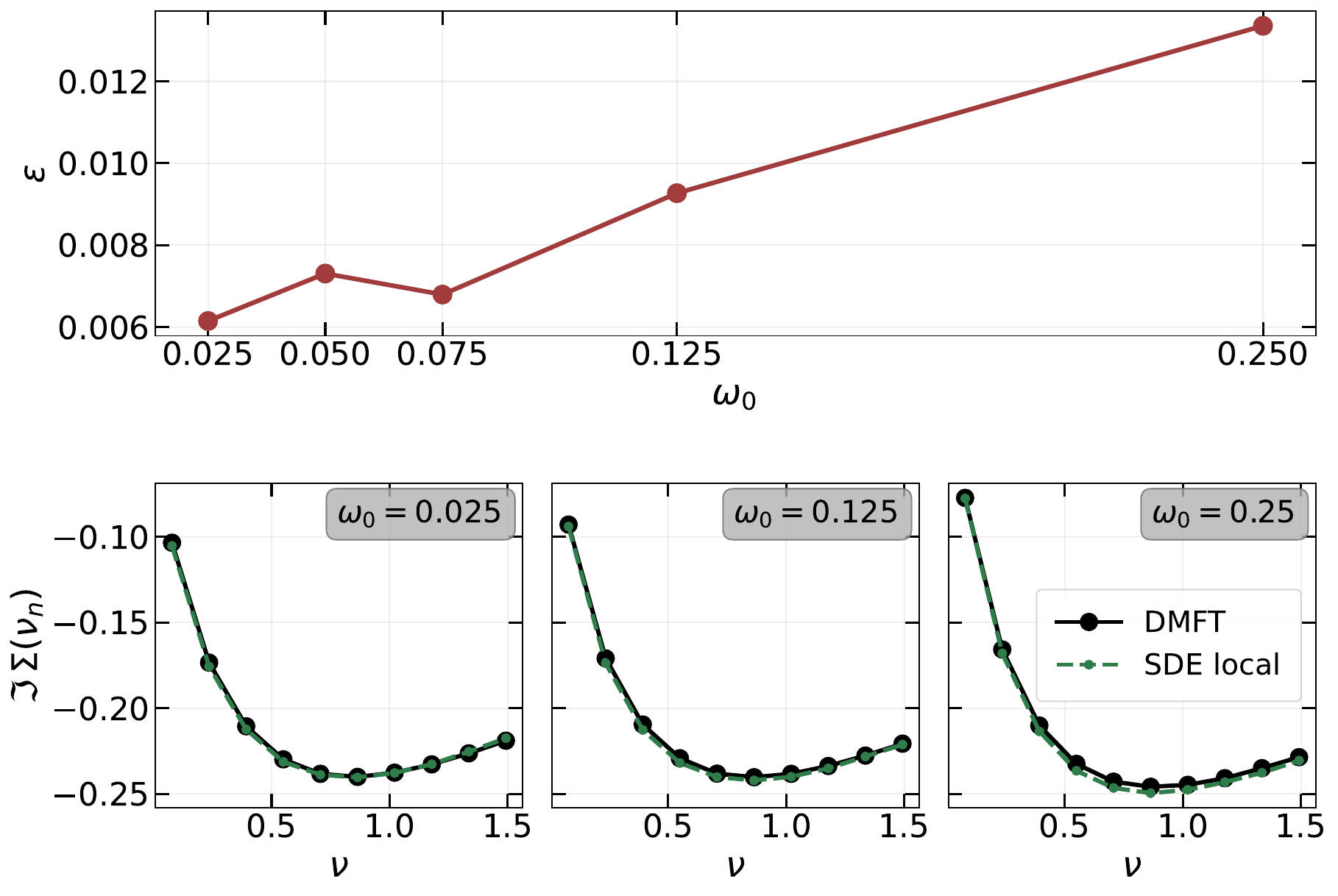}
    \caption{Relative Frobenius error between the local SDE and DMFT self-energies versus phonon frequency $\omega_0$ at $\beta=40$ (top), with representative comparisons of $\Im\Sigma(\nu_n)$ for $\omega_0=0.025$, $0.125$, and $0.25$ (bottom).}
    \label{fig:relerr_dmft_sde}
\end{figure}

In Fig.~\ref{fig:relerr_dmft_sde} we showcase this feature for the Hubbard-Holstein model at a relatively high temperature where errors arising from the frequency box can be clearly distinguished from errors originating from too large values of $\omega_0$. In the upper panel we plot the relative Frobenius error of the local SDE compared to the DMFT self-energy at the first ten positive fermionic Matsubara frequencies for various characteristic phonon scales $\omega_0$ at $\beta=40$ and the parameter regime defined in Sec.~\ref{sec:sc}. The lower panels showcase three examples of the local SDE and DMFT self-energy at $\omega_0={0.025, 0.125, 0.25}$. In both cases we clearly observe a monotonic increase of the error. Lastly, we want to stress here again that this is an intrinsic limitation of the Moriya-$\lambda$-correction approach and can be avoided by employing the self-consistent formulation of ladder-D$\Gamma$A, which does not require a formulation of the SDE in terms of longitudinal interactions in the first place. 

\subsection{Asymptotic treatment of the DMFT vertices}\label{app:asymp}

One of the bottlenecks in diagrammatic extensions of DMFT is undoubtedly the vertex function, which is the main building block of the ladder equations. The dependence of this object on three frequencies renders the numerical treatment one of the main hurdles, as the convergence with the number of Matsubara points in frequency summations or inversions in the BSE can become quite cumbersome. In order to circumvent this issue without sampling the vertices on immense frequency boxes, which can easily become too costly, several sophisticated compression methods have been proposed in the past, which, among other things, facilitate the measurement on smaller, more manageable grids \cite{kunes2011,shinaoka2017,tagliavini2018,krien2019,wentzell2020, shinaoka2023,moghadas2024,rohshap2025}. For our study, we opted for a simplified version of Ref.~\cite{kunes2011} that was introduced in \cite{kitatani2022} and adapted it to the case of retarded interactions.

For the inversion of the BSE we separate the fermionic-frequency space, for every fixed transfer
frequency $\omega$, into a core box $C$, an outer box $O$, and an asymptotic
region $A$, with the frequency-boxes having a size of $n_X$, $X\in\{C,O,X\}$ along each direction. The full impurity vertex is treated explicitly in $C$, whereas in
$O\setminus C$ only its high-frequency contribution obtained from the knowledge of the bare interaction and the one-particle Green's function is retained. In practice $n_O$ can be chosen sufficiently large with $\sim\mathcal{O}(500-1000)$ such that the Green's function's behavior is already governed by its high-frequency $1/i\nu$-tail. Hence the contribution of region $A$ can be determined analytically.

For the Hubbard model this construction separates the constant bare
interaction $U_{d/m} = \pm U$ in the respective channels from the corresponding irreducible vertices. In the present case, the same separation has to be performed with respect to the longitudinal
interaction defined in Eq.~\eqref{equ:longitudinal_interactions}:
\begin{equation}\label{equ:gamma_asymptotic_split}
    \Gamma_{r,\mathrm{loc}}^{\nu\nu'\omega}
    =\mathcal{U}_r^\omega
    +\delta\Gamma_{r,\mathrm{loc}}^{\nu\nu'\omega}
\end{equation}
That is, interaction reducibility is defined by cutting a longitudinal
$\mathcal{U}_r^\omega$ line, rather than a purely instantaneous $U_r$ line.
The transverse part of the generalized bare interaction, which depends on
$\nu-\nu'$, is not used to define this decomposition and remains part of the
residual vertex $\delta\Gamma_r$. Since $\mathcal{U}_r^\omega$ depends only on
the fixed transfer frequency, it is constant in the fermionic-frequency
matrix, exactly as $U_r$ is in the Hubbard model.

We will now summarize the main steps for the asymptotic treatment of the vertex, but refer to \cite{kitatani2022} for the details, as conceptually the mere difference here is replacing the bare $U_r$ with the frequency-dependent $\mathcal{U}_r^\omega$. Throughout the following we will omit frequency-superscripts for the sake of better readability and will denote the longitudinal interaction, which is a matrix in fermionic frequency-space as ${\left[\mathcal{U}_{r,X}^\omega \right]}_{\nu\nu'}=\mathcal{U}_r$. Also, since for a non-diagonal matrix, truncation and inversion do not commute, we refer to the operation $(A_O)^{-1}_C$ as the inversion of the matrix $A$ on the $O$-region and then truncating it down to its $C$-region. First, we define the local BSE that we want to solve numerically on the $O$-region grid as
\begin{equation}\label{equ:loc_bse_o}
    \left(\chi_O^{loc}\right)^{-1} = \left(\chi_{0,O}^{loc}\right)^{-1} + \left( \mathcal{U}_{O}  + \delta \Gamma_{C}\right).
\end{equation}
Subsequently, defining $\left(\chi_{0,O}^{loc}\right)^{-1} + \mathcal{U}_{O} = \left(\widetilde{\chi}_{O}^{loc}\right)^{-1}$, we can solve for the irreducible vertex via
\begin{equation}
    \delta\Gamma_C = \left(\chi_O^{loc}\right)^{-1}_C - \left(\widetilde{\chi}_{O}^{loc}\right)^{-1}
\end{equation}
While we do not know $\chi^{loc}$ in the $O$-region, by using the fact that $\delta\Gamma$ vanishes in the $O$-region, we can still solve the above BSE
\begin{equation}
    \left(\chi_O^{loc}\right)_C = \left(\widetilde{\chi}_{O}^{loc}\right)_C - \sum_{\nu,\nu'\in C} \widetilde{\chi}_{O}^{loc} \, \delta \Gamma_c \, \chi_O^{loc}
\end{equation}
where the summation over internal frequency variables is performed only in the $C$-region. Inverting this equation we obtain an expression for $\delta\Gamma$ containing information on the high-frequency asymptotics, where it is sufficient to know $\chi^{loc}$ only in the $C$-region
\begin{equation}\label{equ:Gamma_c_final}
    \delta\Gamma_C = \left(\chi_C^{loc}\right)^{-1} - \left( \left(\widetilde{\chi}_{O}^{loc}\right)_C \right)^{-1}
\end{equation}

\begin{figure}[h!]
    \centering
    \includegraphics[width=1.0\linewidth]{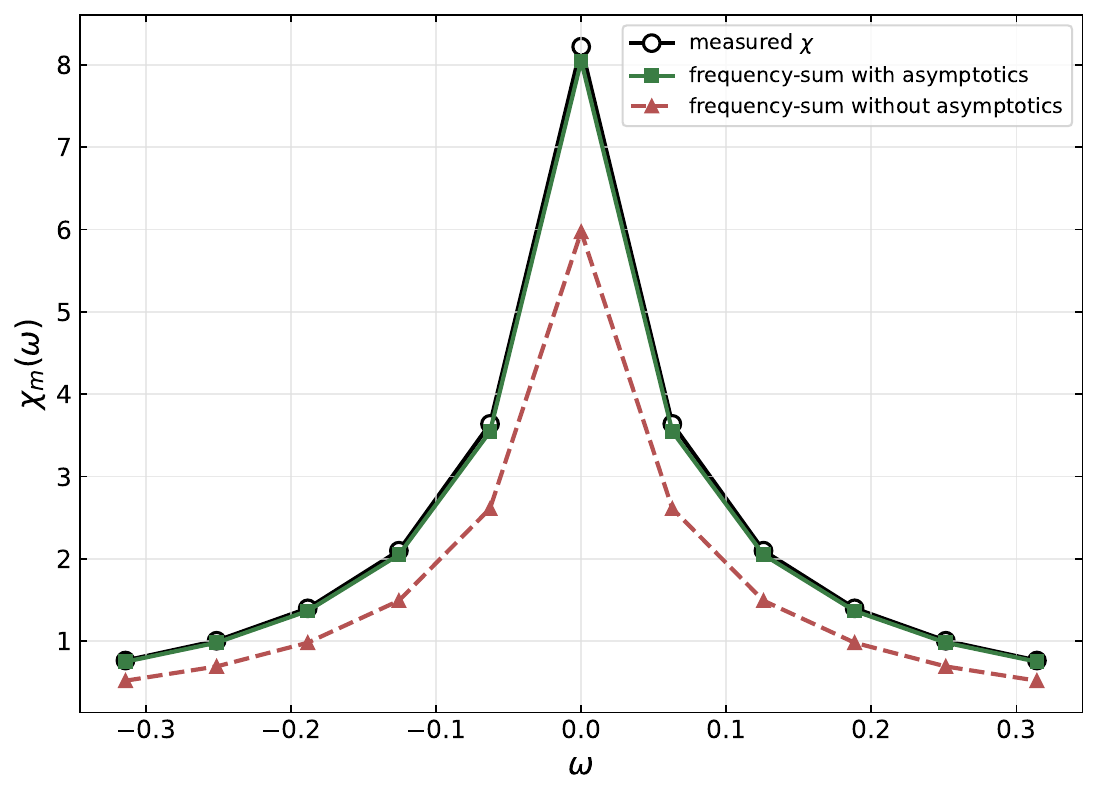}
    \caption{Local magnetic susceptibility of the Hubbard--Holstein model at
    $\beta=100$, $n=0.8$, $\Omega_0=0.1$, $U=3$, and $\lambda=1$, calculated
    with $n_{\mathrm{inner}}=100$ and $n_{\mathrm{outer}}=1500$. The direct
    impurity measurement is compared with Matsubara-frequency sums with and
    without the asymptotic correction.}
    \label{fig:chi_asymp}
\end{figure}

Utilizing Eq.~\eqref{equ:Gamma_c_final} allows us to obtain expressions for the asymptotics of the non-local quantities of the lattice \cite{kitatani2022}. These are the physical lattice response function $\chi_{r,\mathbf{q}}$ and Hedin vertex $\gamma_{r,\mathbf{q}}$ given in Eq.~\eqref{equ:chi_gamma_q} and the interaction-irreducible generalized susceptibility $\chi^*_{r,\mathbf{q}}$ of Eq.~\eqref{equ:chi_star}. We present here, for the sake of completeness, the asymptotic expressions for the lattice quantities, originally derived in Ref.~\cite{kitatani2022}, where the overline denotes a sum over fermionic frequencies, $\overline{\chi_X}\equiv\sum_{\nu\nu'}\chi_X$ and lattice quantities are marked with a \textit{latt} superscript.

The interaction-irreducible susceptibility in the $C$-region can be then obtained with the improved $\delta\Gamma_C$ of Eq.\eqref{equ:Gamma_c_final} via
\begin{equation}\label{equ:chi_star_U}
    \chi^{*,latt}_C = \left( \left( \chi^{latt}_{0,C} \right)^{-1} + \delta\Gamma_C \right)^{-1}
\end{equation}
From this we compute the following asymptotic expression for the physical lattice response function
\begin{equation}\label{equ_chi_phys_asymp}
    \begin{aligned}
        {}& \chi^{latt}_{asymp} \\
        &= \left( \left( \overline{\chi^{*,latt}_C} + \overline{\chi^{latt}_{0,O}} + \overline{\chi^{latt}_{0,C}}\right)^{-1} + \mathcal{U}_r \right)^{-1} + \overline{\chi^{latt}_{0,A}}-\overline{\chi^{latt}_{0,O}}
    \end{aligned}
\end{equation}
where the difference in the last two terms corresponds to the frequency-sum over the asymptotic tail of the bubble term, $\chi_0 \sim \frac{\delta_{\nu\nu'}}{\nu^2}$, which can be evaluated analytically \cite{reitner2020attractive}
\begin{equation}
    \begin{aligned}
        \chi_{0,asymp} {}&\equiv \overline{\chi^{latt}_{0,A}}-\overline{\chi^{latt}_{0,O}} \\
        &= \frac{1}{\beta}\sum_{n_O}^{\infty} \frac{1}{\nu^2} = \frac{\beta}{\pi^2}\left( \frac{\pi^2}{8} - \sum_{\tilde{n}=0}^{n_O} \frac{1}{(2\tilde{n}+1)^2}\right)
    \end{aligned}
\end{equation}
The correction for the three-leg vertex is given by
\begin{equation}
    (\gamma^{latt}_{asymp})_C = \left(\chi^{latt}_{0,C}\right)^{-1} \chi_C^{*,latt} \frac{1-\mathcal{U}_r(\chi^{latt}_{asymp}-\chi_{0,asymp})}{1-\mathcal{U}_r \chi^{latt}_{asymp}}.
\end{equation}
Eventually, the asymptotic correction for the interaction-irreducible susceptibility can be computed as
\begin{equation}
    \begin{aligned}
    {}& \left(\chi^{*,latt}_{asymp}\right)_C \\
    &= \chi^{*,latt}_C - \chi^{*,latt}_C \mathcal{U}_r \chi^{*,latt}_C + \\
    &+ \chi^{*,latt}_C \mathcal{U}_r \chi^{*,latt}_C \frac{\left(1-\mathcal{U}_r(\chi^{latt}_{asymp}-\chi_{0,asymp})\right)^2}{1-\mathcal{U}_r \chi^{latt}_{asymp}}
    \end{aligned}
\end{equation}

As a concrete example, Fig.~\ref{fig:chi_asymp} shows the local magnetic
susceptibility of the Hubbard-Holstein model at $\beta=100$, $n=0.8$,
$\Omega_0=0.1$, $U=3$, and $\lambda=1$. Here the directly evaluated inner
box corresponds to the core, with $n_{\mathrm{inner}}=100$, while the
longitudinal-interaction asymptotics is included up to
$n_{\mathrm{outer}}=1500$. Supplementing the outer-frequency contribution
recovers the susceptibility measured directly by the impurity solver, whereas
the frequency sum restricted to the inner box substantially underestimates
it, particularly at small $\omega_n$.

\subsection{Fitting details}\label{app:fit}
\begin{figure*}
    \centering
    \includegraphics[width=1.0\linewidth]{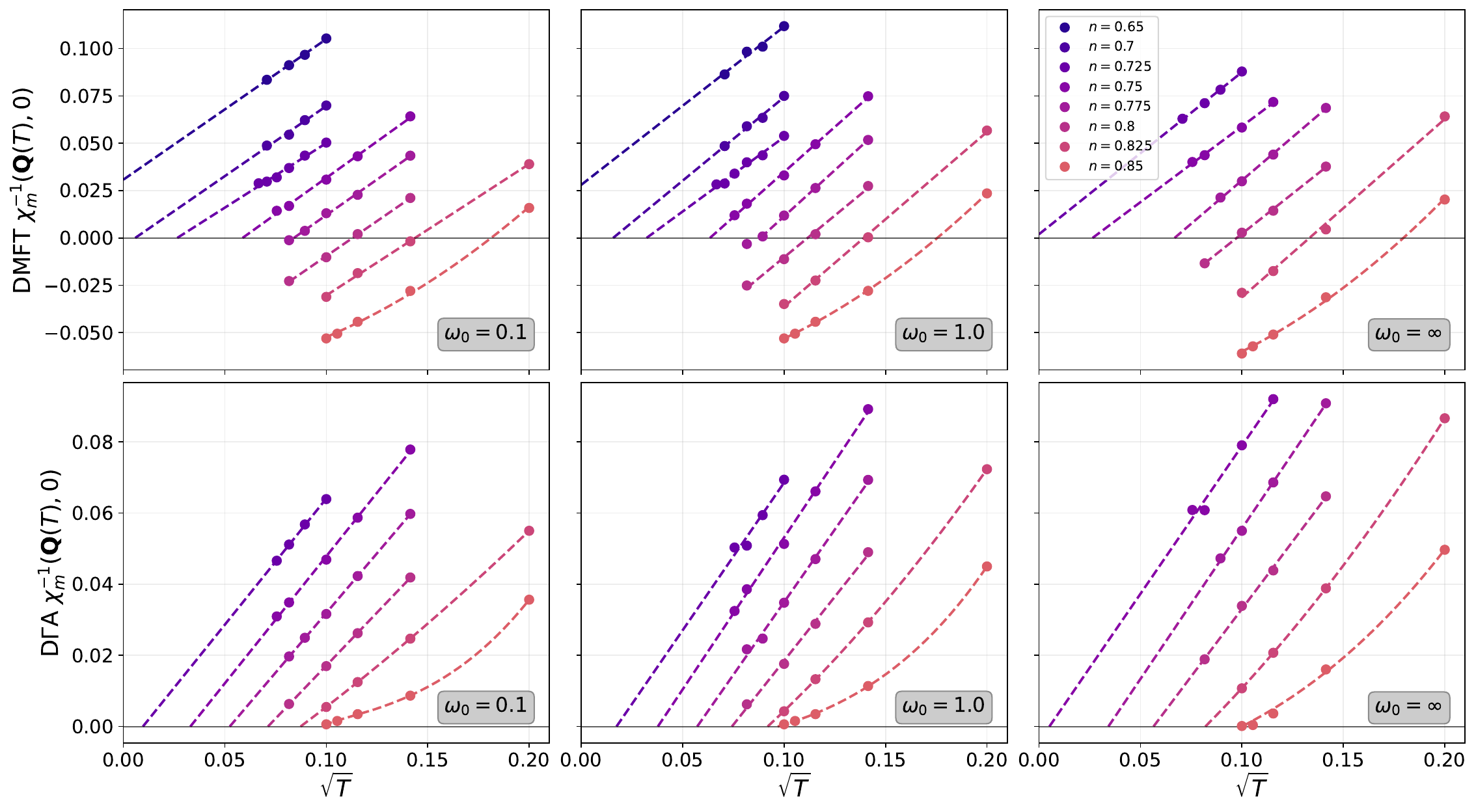}
    \caption{Temperature dependence of the DMFT (upper row) and D$\Gamma$A (lower row) inverse magnetic susceptibilities of the Hubbard-Holstein model in their static limits at the ordering vector $\mathbf{Q}$ for fillings below $n\leq0.85$ and the three phonon frequencies $\omega_0=\{0.1,1,\infty \}$. For all calculations considered here the interactions are set to $U=3$ and $\lambda=1$.}
    \label{fig:chi_fit_zoom}
\end{figure*}
In this section we provide further details on the temperature fits to the $\chi^{-1}$ data for Sec.~\ref{sec:afm}. For the rather straightforward case of half-filling, throughout all values of $\lambda$ and $\omega_0$ the same fitting procedure to the four lowest temperature data points was employed in DMFT and D$\Gamma$A. For both cases, the lowest four temperature data points were used for the fits. In DMFT, where the physical response function behaves like $\chi_m\propto T^{-\gamma}$, with $\gamma=1$, the transition temperature was obtained from the roots of a linear fit to the data. As mentioned in the main text, the TPSC universality class would yield, in the 3D case, $\gamma=2$ for Moriya-corrected ladder-D$\Gamma$A. In this context, we found that adding a subleading cubic term \cite{delre2019} resulted in improved estimates of $T_N$. 

Once the system is doped, as discussed in the main text, the Kohn anomalies significantly change the quantum (classical) critical behavior requiring a different fitting function than for half-filling. Following Ref.~\cite{fus2026} for densities close to the QCP, i.e., $n\in{0.65,0,825}$, we fit the data according to $\chi_m^{-1} \propto a + b\sqrt{T}$ to determine the transition temperature. For densities closer to half-filling we used the regular, linear and quadratic forms for DMFT and D$\Gamma$A, respectively. The sole exception here is the case of the D$\Gamma$A data of $n=0.85$. In this particular regime between Kohn-anomalies driven quantum-criticality and the critical behavior dictated by the Moriya-theory, we found that more subleading terms according to $\chi_m^{-1} \propto a + b\sqrt{T} + cT + dT^2$ yield the most optimal fit to the data. As an example, in Fig.~\ref{fig:chi_fit_zoom} we provide again the fit to the inverse susceptibilities already reported in the main text, now only for the lowest densities and against a $\sqrt{T}$ x-axis to highlight their temperature dependence close to the QCP. We clearly observe the square-root behavior of the DMFT (D$\Gamma$A) response functions that becomes linear (quadratic) around $n\sim0.85$.

\subsection{Comparison with a $d$-wave BCS model}\label{app:bcs_comparison}

To place the superconducting eigenvectors of the Hubbard--Holstein model in context, we compare them with those of a simple BCS-like model. We choose a separable $d$-wave interaction because previous D$\Gamma$A studies have established this symmetry as the leading superconducting instability in the doped Hubbard model \cite{kitatani2019,kitatani2022,rohringer2018diagrammatic}. Using the same gap symmetry allows us to compare the momentum and frequency dependence of the leading Eliashberg eigenvectors and identify their characteristic energy scales. The model interaction is defined as
\begin{equation}\label{equ:bsc_V}
    \Gamma_{BCS}^{kk'} = V f(\mathbf{k}) f(\mathbf{k'}) C(k) C(k')
\end{equation}
with $f(\mathbf{k}) = \cos(k_x) - \cos(k_y)$ being a $d$-wave form-factor and $C(k)$ denoting a cutoff function in the fermionic-four-momentum depending on the Debye-frequency $\omega_D$. To mimic finite temperature blurring, w.l.o.g. we choose the latter to be a \textit{broadened} Heaviside function $C(k) \equiv c_{\delta}(\mathbf{k})c_{\delta}(\nu)$ defined as
\begin{equation}
    c_{\delta}(x) = \left[1 + \exp \left(\frac{|x|-\omega_D}{\delta}\right) \right]^{-1}
\end{equation}
where in the limit $\delta \to 0$ the regular step function is recovered. Plugging Eq.~\eqref{equ:bsc_V} into the linearized Eliashberg equation of Eq.~\eqref{equ:eliashberg} will yield
\begin{equation}
    \begin{aligned}
        &{} \epsilon_{BCS} \, \Delta(k) \\
        &{} = -\frac{V f(\mathbf{k}) C(k)}{\beta}\sum_{k'} f(\mathbf{k'}) C(k')G(k')G(-k')\Delta(k')
    \end{aligned}
\end{equation}
Evidently, the entire $k$-dependence of the gap-function originates from $f(\mathbf{k}) C(k)$, and hence will inherit the characteristic scale given by $\omega_D$.

As we will see, in contrast to the BCS case, the Hubbard-Holstein model does not feature an intrinsic phonon scale in the pairing vertex: the scale for the gap-function is set by the electronic bandwidth.

\begin{figure*}
    \centering
    \includegraphics[width=1.0\linewidth]{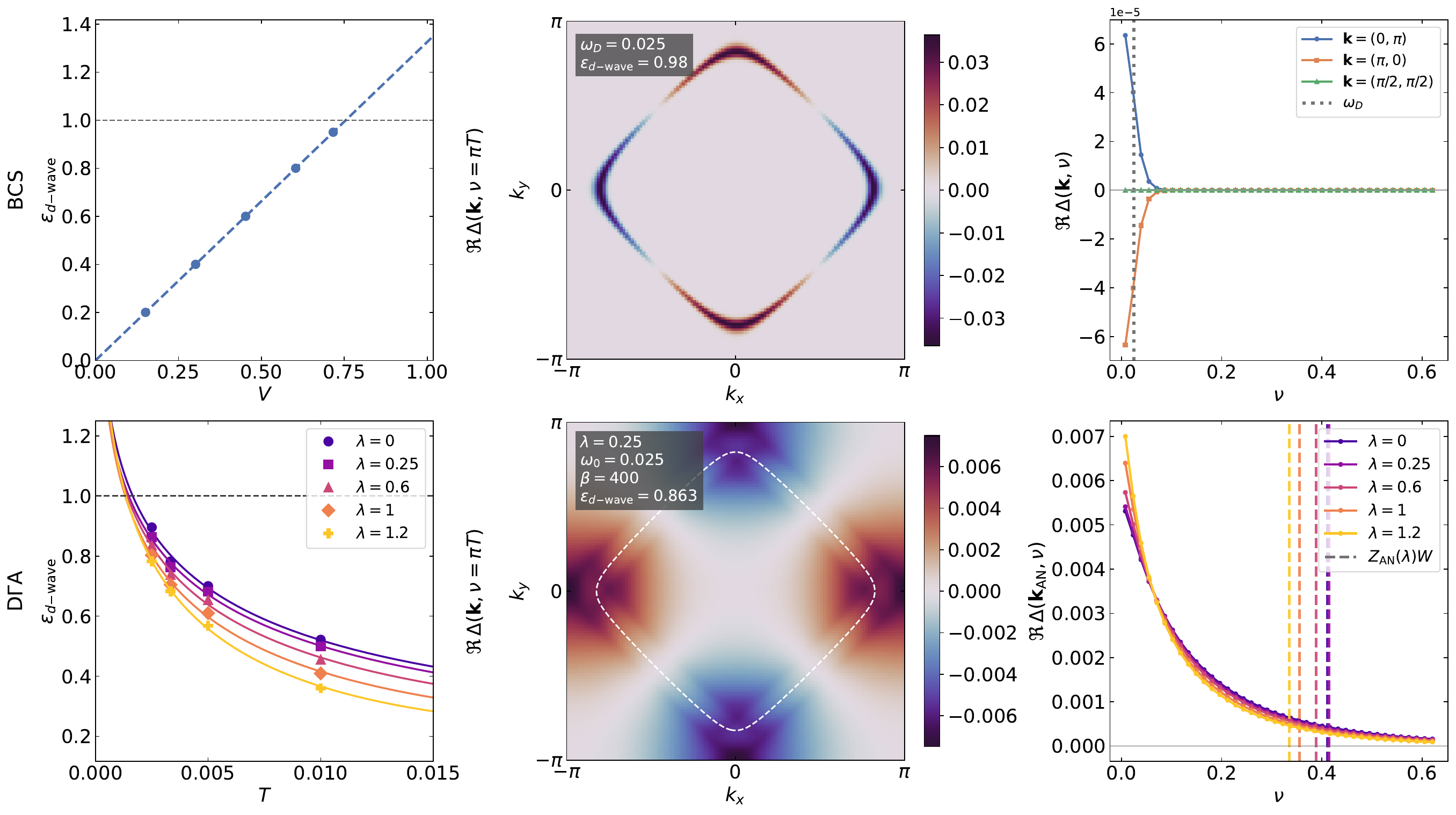}
    \caption{Upper row: Example of the eigenvalues and eigenvectors of the linearized Eliashberg equation for the $d$-wave BCS-model evaluated at $\beta=100$, $\omega_D=0.025$ and $n=0.8$. In the left panel the evolution of the eigenvalues with the bare interaction is plotted. The middle panel shows the momentum-dependence of the real part of the leading eigenvectors for the smallest fermionic frequencies $\nu=\pi T$ close to the superconducting transition. The right panel compares the frequency-dependence of the real part of the leading eigenvector for select momenta on the Fermi-surface with the characteristic frequency $\omega_D$. Lower row: Example of the eigenvalues and eigenvectors of the linearized Eliashberg equation for the Hubbard-Holstein model at $\omega_0=0.025$ and $n=0.825$. In the left panel the temperature-dependence of the leading eigenvalues for various $\lambda$ is plotted. The middle panel showcases the momentum-dependence of the real part of the leading eigenvectors for the smallest fermionic frequencies $\nu=\pi T$ close to the transition, specifically at $\beta=400$, $\lambda=0.25$ with the white dashed line denoting the Fermi-surface. In the rightmost panel the frequency-dependence of the real part of the leading eigenvector at the antinodal point is compared for different values of $\lambda$. The vertical dashed lines denote the renormalized electronic bandwidth.}
    \label{fig:dga_bcs}
\end{figure*}

The upper row of Fig.~\ref{fig:dga_bcs} shows a representative calculation for the $d$-wave BCS model at $n=0.8$, $\beta=100$, and $\omega_D=0.025$. There, increasing the pairing interaction $V$ drives the leading eigenvalue through unity in the left panel, while the middle and right panels show the momentum and frequency dependence of the associated eigenvector, respectively. In the BCS case, the frequency dependence is restricted by the Debye scale $\omega_D$. For comparison, the lower row demonstrates representative eigenvalues and eigenvectors of the linearized gap equation for the Hubbard--Holstein model at $\omega_0=0.025$ obtained within our D$\Gamma$A approach. The left panel shows the temperature dependence of the leading $d$-wave eigenvalue $\epsilon_{d\text{-wave}}$ for several el-ph couplings $\lambda$; its crossing of unity, indicated by the horizontal dashed line, determines the superconducting transition temperature $T_c$. The middle panel displays the momentum dependence of the corresponding gap function at the lowest fermionic Matsubara frequency, $\nu=\pi T$, for the parameter set specified in the gray box near the transition. The white dashed contour marks the interacting FS obtained from the D$\Gamma$A self-energy. The right panel shows the dependence of the gap function on Matsubara frequency at the antinodal momentum $\mathbf{k}_{\mathrm{AN}}$ and the lowest available temperature for different values of $\lambda$. The lines indicate the corresponding renormalized electronic scale $Z(\lambda,\mathbf{k}_{\mathrm{AN}})W$. Unlike the BCS result, the D$\Gamma$A gap function extends over a broader region of momentum and frequency space and correlates with the renormalized electronic bandwidth. This qualitative distinction directly illustrates that phonons do not provide a conventional BCS pairing-glue in the D$\Gamma$A calculation, but instead modify a correlation-driven $d$-wave instability whose characteristic scale remains predominantly electronic.

\section*{Data Availability}
A data set containing all numerical data and plot scripts used to generate the figures of this publication is publicly available at \cite{note:data-repo}. The ladder-D$\Gamma$A code is publicly available under \cite{moghadas_ldga}.


\bibliography{bib/library} 

\end{document}